\documentclass[sigconf]{acmart}
\AtBeginDocument{%
  }

\copyrightyear{2026}
\acmYear{2026}
\setcopyright{cc}
\setcctype{by}
\acmConference[CCS '26]{Proceedings of the 2026 ACM SIGSAC Conference on Computer and Communications Security}{November 15--19, 2026}{The Hague, Netherlands}
\acmBooktitle{Proceedings of the 2026 ACM SIGSAC Conference on Computer and Communications Security (CCS '26), November 15--19, 2026, The Hague, Netherlands}
\acmDOI{10.1145/3830454.3846693}
\acmISBN{979-8-4007-2871-6/2026/11}

\usepackage{tikz}
\usepackage{amsmath}
\usepackage{csquotes}
\usepackage{enumitem}

\usepackage{pifont}
\usepackage{graphicx}
\usepackage{booktabs}
\usepackage[most]{tcolorbox}
\usepackage{cleveref}
\usepackage{glossaries}
\makeglossaries
\glsdisablehyper
\newacronym{us}{U.S.}{United States}
\newacronym{uscg}{USCG}{United States Coast Guard}
\newacronym{usmc}{USMC}{U.S. Marine Corps}
\newacronym{cg}{CG}{Coast Guard}
\newacronym{usn}{USN}{United States Navy}
\newacronym[longplural={Officers of the Deck}]{ood}{OOD}{Officer of the Deck}
\newacronym{dwo}{DWO}{Deck Watch Officer}
\newacronym{swo}{SWO}{Surface Warfare Officer}
\newacronym{dod}{DoD}{Department of Defense}
\newacronym{gps}{GPS}{Global Positioning System}
\newacronym{cps}{CPS}{cyber-physical systems}
\newacronym{flss}{FLSS}{Fire Life-Safety Systems}
\newacronym{IT}{IT}{Information Systems Technician}
\newacronym{ET}{ET}{Electronics Technician}
\newacronym{IRB}{IRB}{Institutional Review Board}
\newacronym{RQ}{RQ}{research question}
\newacronym{infotech}{IT}{information technology}
\newacronym{OT}{OT}{operational technology}
\newacronym{OSF}{OSF}{Open Science Framework}
\newacronym{ENAV}{ENAV}{Electronic Navigation}
\newacronym{CO}{CO}{Commanding Officer}
\newacronym{DC}{DC}{Damage Control}
\newacronym{OPSEC}{OPSEC}{Operational Security}
\newacronym{EMCON}{EMCON}{Emission Control}
\newacronym{NAV}{NAV}{Navigator}
\newacronym{OPS}{OPS}{Operations Officer}
\newacronym{CSO}{CSO}{Command Security Officer}
\newacronym{CYBO}{CYBO}{Cyber Officer}
\newacronym{AIS}{AIS}{Automatic Identification System}
\newacronym{ATFP}{ATFP}{Antiterrorism and Force Protection}

\crefname{enumi}{RQ}{RQs}
\Crefname{enumi}{RQ}{RQs}

\begin{document}

\title{Batten the Hatches: Cybersecurity with Military Mariners}

\author{Ryan Von Brock}
\orcid{0009-0000-7093-4667}
\affiliation{
  \institution{Georgia Institute of Technology}
  \city{Atlanta}
  \state{GA}
  \country{USA}
}
\email{vonbrock@gatech.edu}

\author{Anna Raymaker}
\orcid{0000-0001-7093-1984}
\affiliation{
  \institution{Georgia Institute of Technology}
  \city{Atlanta}
  \state{GA}
  \country{USA}
}
\email{araymaker3@gatech.edu}

\author{Animesh Chhotaray}
\orcid{0009-0006-1051-172X}
\affiliation{
  \institution{Georgia Institute of Technology}
  \city{Atlanta}
  \state{GA}
  \country{USA}
}
\email{achhotaray3@gatech.edu}

\author{Frank Li}
\orcid{0000-0003-2242-048X}
\affiliation{
  \institution{Georgia Institute of Technology}
  \city{Atlanta}
  \state{GA}
  \country{USA}
}
\email{frankli@gatech.edu}

\author{Saman Zonouz}
\orcid{0009-0006-7302-0178}
\affiliation{
  \institution{Georgia Institute of Technology}
  \city{Atlanta}
  \state{GA}
  \country{USA}
}
\email{szonouz6@gatech.edu}

\author{Raheem Beyah}
\orcid{0000-0002-9188-3464}
\affiliation{
  \institution{Georgia Institute of Technology}
  \city{Atlanta}
  \state{GA}
  \country{USA}
}
\email{ab207@gatech.edu}

\renewcommand{\shortauthors}{Ryan Von Brock et al.}

\begin{abstract}

Cyberwarfare has become a key component of contemporary geopolitical conflict. However, there has been extremely limited systematic investigation into how cybersecurity is handled by military organizations and personnel. The military context is unique compared to other operational ones, with immense resource availability (U.S. military spending approached 1 trillion dollars in 2024), a rigid chain of command, and extraordinary consequences for its actions. Thus, military cybersecurity is a distinct yet understudied topic.

In this paper, we take an early step at understanding military cybersecurity by investigating how service members understand, recognize, and respond to cyber risk. We focus on maritime services and carefully consider organizational barriers to design an unclassified study and conduct semi-structured interviews with 20 military mariners from U.S. Navy and Coast Guard vessels. Through our investigation, we identify unique consequences of compromising military systems, including weapon takeover and purposeful geopolitical escalation. We find that cybersecurity is organizationally abstract on ships, so mariners build cyber risk models from informal experience rather than formal instruction. They nonetheless make cybersecurity actionable by recognizing operational impacts and responding with a safety-oriented incident-response model that creates resilience but may delay cyber attribution and containment. These findings inform actionable recommendations to help military operators frame cyber threats, merge longstanding nautical doctrine with modern systems, and apply military insights to the civilian sector, all to secure the broader maritime environment.

\end{abstract}



\maketitle

\section{Introduction}
\label{sec:intro}

Cyberwarfare has become a central component of modern geopolitical conflict, with cyber operations increasingly used to gain battlefield advantages~\cite{ferazza2025non, ViasatOverview, ViasatCaseStudy}. In military environments, cybersecurity failures can escalate beyond financial harm to cause mission degradation, loss of life, and strategic consequences~\cite{reddy2025cyber, lonergan2023power}. Unlike civilian sectors driven by market incentives, military organizations optimize for capability, resilience, and readiness~\cite{hartley2012economics}.

Despite sustained investment in securing military platforms, scale, heterogeneity, and mission demands introduce persistent challenges~\cite{koch2016weapons}. Modern naval systems are tightly coupled cyber-physical environments in which digital failures directly influence physical outcomes. Prior work has extensively studied technical vulnerabilities and defenses in \gls{cps}, including the automotive~\cite{xue2022said, wen2020plug, jing2024revisiting, hu2021automated, bhatia2021evading}, aviation~\cite{jansen2021trust, jansen2017localization, birnbach2017wi, lundberg2014security}, and industrial control sectors~\cite{pickren2024release, singer2023shedding, sasaki2022exposed, abukhousa2025wisdom}, yet empirical research on the human factors of military cybersecurity remains scarce. Many service members who operate mission-critical systems are not cybersecurity specialists, but their decisions shape real-world security outcomes. Recent incidents, including the unauthorized installation of satellite communication equipment aboard a navy combat ship and misconfiguration of military email servers, demonstrate that human actions can bypass formal organizational safeguards~\cite{navy_wifi, dod_email_leak, dod_email_domain}. These cases suggest that improving military cybersecurity requires analyzing operator behavior.

The military sector remains largely absent from unclassified empirical cybersecurity research. Existing studies emphasize doctrine, high-level strategy, or system design~\cite{ferazza2025non, bottero2025systems}, providing limited visibility into how cybersecurity is enacted by operators under real constraints. The small body of peer-reviewed empirical work on military cybersecurity focuses on organizational threat-hunting processes rather than the perceptions and decisions of frontline personnel~\cite{maxam2024interview}. Access restrictions and cultural barriers have contributed to this gap, leaving open questions about how policy and training translate into operational practice.

In this paper, we present an unclassified qualitative study of military mariners to address this gap. We conducted semi-structured interviews with 20 \gls{usn} and \gls{uscg} mariners, asking perception questions and presenting three cyber-physical scenarios to study their operational reasoning and investigate the following \glspl{RQ}:
\begin{enumerate}[label=\textbf{RQ\arabic*:}, ref=\arabic*, leftmargin=*]
  \item \label{rq:1} What factors shape \gls{us} military mariners' mental models of cybersecurity risk? 
  \item \label{rq:2} How do \gls{us} military mariners recognize and respond to cybersecurity threats under organizational constraints? 
\end{enumerate}

\crefformat{enumi}{#2RQ#1#3}
\crefmultiformat{enumi}{RQ#2#1#3}{ and RQ#2#1#3}{, RQ#2#1#3}{, and RQ#2#1#3}

We focus on maritime services as a subset of the military for several reasons. Their global deployments and multinational training environments create shared operational practices and allow for comparison with civilian maritime contexts. Secondly, naval warships and \gls{cg} cutters are among the most complex self-contained \gls{cps} in operation, integrating navigation, propulsion, weapons, and communication systems into interdependent architectures~\cite{bottero2025systems, progoulakis2021cyber}. Lastly, their strategic roles amplify the consequences of cyber failure: navies underpin deterrence and force projection, while \gls{cg} missions span military operations, law enforcement, and the regulation of commercial shipping~\cite{cg_mission, navy_mission}. \gls{cg} personnel bridge military and civilian maritime domains. Understanding how military mariners operationalize cybersecurity informs both military and broader maritime security.

Our analysis yields two overarching findings. First, cybersecurity is organizationally abstract for military mariners: deferred in priority, ambiguous in responsibility, and framed by situational context rather than doctrine. Mariners therefore build their models of cyber risk from informal experience and situational cues, reflecting a structural mismatch between \gls{infotech}-centric, compliance-driven training and the realities of shipboard operations. This also shapes their focus, as participants' threat models often center on disruptions to navigation systems, while the most severe risks they describe involve integrity breaches. Second, mariners make cybersecurity actionable by recognizing operational impacts and responding with practiced seamanship. This safety-oriented incident response model, optimized for physical casualties, creates resilience but can delay cyber attribution and containment. Together, these findings reveal a disconnect between how cyber incidents are framed and how operators make decisions in practice, leading to blind spots that technical safeguards alone do not address. We also give a comparative analysis of military and civilian mariner cybersecurity perspectives, revealing significant disparities in threat awareness, perceived impact, and mitigation complexity. Based on this analysis, we offer actionable recommendations to enhance maritime security.

In summary, we make the following contributions in this paper:
\begin{itemize}[leftmargin=*,noitemsep]
    \item We derive a model of how military mariners detect, interpret, and respond to cyber-physical incidents.
    \item We show that the organizational abstraction of cybersecurity leaves operator heuristics misaligned with integrity-focused attacks, explaining why certain threats may evade timely response.
    \item We provide recommendations to enhance maritime safety by comparing military and civilian mariner perspectives and identifying cross-domain lessons for the broader maritime ecosystem.
\end{itemize}

\section{Background and Related Work}
\label{sec:background}

In this section, we specify the study's target audience, define key terms, and highlight cybersecurity challenges for ships. We also review existing research across similar domains.

\subsection{Background}
We define the term \textit{military mariners} as the personnel who operate systems aboard \gls{cg} and navy vessels. Our study focuses on the experiences of \glspl{ood}, but we augment their experiences with engineers' and other technical experts'. \glspl{ood} are the designated officers in charge of and responsible for an entire naval vessel's operation~\cite{uscg_ood}. Their decision-making and oversight of shipboard duties put them in a unique position to observe and respond to cyber-related risks during military operations. Appendix~\ref{app:military_defs} provides a brief description of military rank and job structure.

\textbf{Military Vessels} are some of the most complex \gls{cps} in operation today~\cite{bottero2025systems, progoulakis2021cyber}. At a minimum, vessels integrate navigation, power generation, propulsion, and life support systems. Military ships also incorporate systems and capabilities that distinguish them from merchant platforms and bring unique cybersecurity considerations. Navigation systems may utilize restricted signals, such as M-code \gls{gps}, alongside advanced inertial navigation systems~\cite{gps_mcode, navy_imu}. Power systems range from conventional diesel engines to nuclear reactors, where monitoring and control systems interface with safety-critical processes~\cite{navy_propulsion}. Weapon systems, including missile launchers, conventional guns, and countermeasures, are tightly integrated with sensors and command networks~\cite{navy_weapons}. The presence of classified subsystems further complicates system integration and incident response.

\textbf{Cyber Risks:} ships are exposed to a range of well-documented cyber and electronic attacks that can affect safety-critical operations. \gls{gps} spoofing and jamming can degrade or falsify positions, leading to navigational errors. The takeover of vessels through \gls{gps} spoofing and defense mechanisms are well investigated by researchers~\cite{bhatti2017hostile, liu2021stars, ranganathan2016spree, sathaye2022semperfi}. Interference with propulsion and steering control systems can occur through system compromise or the manipulation of sensor inputs. It can impair maneuverability and lead to collision or grounding. Communication jamming or spoofing can isolate vessels or falsify
messages. Similarly, sensor attacks on radar or situational awareness systems can obscure nearby traffic or threats. Compromised \gls{flss} can trigger equipment shutdowns or create unsafe situations, and ransomware can be used to corrupt systems and halt maritime operations~\cite{bimco2024risk, imo2025risk, imo2017}. These attacks vary in sophistication, but they all undermine the integrity and availability of critical systems.

Military systems are assessed for vulnerabilities and designed to mitigate such attacks. For example, encrypted \gls{gps} and inertial navigation systems provide spoofing-resilient navigation, and secure communication systems provide confidentiality and integrity for operational commands~\cite{gps_mcode, navy_imu}. Even with these safeguards in place, the operator must recognize anomalies and transition between systems. Strengthening military cybersecurity requires better understanding the human perspective of shipboard risks. 

\subsection{Related Work}
Unclassified \textit{military cybersecurity research} remains extremely limited, particularly work that examines the operator perspective. A cyber threat hunting study of the \gls{us} Department of Homeland Security documented institutional approaches to identifying cyber threats but focused on organizational processes~\cite{maxam2024interview}. Other studies have provided strategic overviews of cyberwarfare while neglecting operator experiences~\cite{ferazza2025non, reddy2025cyber}. One maritime dissertation investigated cybersecurity awareness within the Ghana Navy; however, this work reflects a developing navy with different resource constraints and missions~\cite{forson2022assessing}. 

In contrast, \textit{civilian-focused cybersecurity} research includes many operator perspectives in sectors like industrial control systems~\cite{fung2025adopting, evripidou2024understanding, evripidou2023exploring, gallardo2024interdisciplinary, li2024usability, flaa2024cybersecurity}, individual security practices~\cite{das2014effect, sahin2023investigating}, and technical defenses~\cite{votipka2018hackers, akgul2023bug, wong2024comparing}. Many of these techniques are transferable to the military, but the operators' perspectives are formed by different resource constraints and organizational factors. Previous work in \textit{maritime cybersecurity} is also comparatively better developed, investigating fine-grained vulnerabilities and defenses in navigation systems~\cite{ranganathan2016spree, liu2021stars, tibaldo2025gnss, sathaye2022semperfi}, satellite communications~\cite{bisping2024wireless, pavur2020tale} or other specialized equipment~\cite{longo2023attacking, tran2021marine}. These vulnerabilities are relevant to some military platforms, but they do not consider human decision-making. Other work characterizes high-level challenges and attack vectors within the maritime domain~\cite{progoulakis2021cyber}, but user-centered maritime research is sparse. Some studies have focused on shore-side maritime terminal operators or high-level policy experts, but these fail to incorporate the underway operator perspective~\cite{nganga2024enabling, fenton2024preventing}. Raymaker et al. explored maritime cyber risk through interviews with merchant and some military-affiliated mariners. This work took an important step toward incorporating practitioner perspectives, but many of the military-affiliated mariners were civilian Military Sealift Command mariners rather than active service members~\cite{raymaker2025sea}. As such, that study reflects a hybrid civilian-military context.

In contrast, this study focuses on the experiences of military mariners operating aboard Navy and \gls{cg} vessels. It is unique in two major ways. First, by exclusively targeting military personnel, we cover a greater breadth of experience and ask military-specific questions, like those addressing rank, military culture, and chain of command. Second, we use scenario-based questions to study mariners' response processing, which yields novel findings. We provide one of the first unclassified, military-focused looks at how mariners perceive, interpret, and respond to cyber incidents at sea.

\section{Methodology} 

We conducted interviews with 20 participants to investigate military cybersecurity practices in the maritime services and address our research questions. The study was approved by our \gls{IRB} and followed standard qualitative cybersecurity research practices, including a pilot study, screening survey, iterative codebook development, multi-coder reliability checks, and saturation-based stopping criteria~\cite{raymaker2025sea, maxam2024interview, das2014effect}. These methods allowed us to identify recurring themes while preserving the nuance of individual participants' experiences.

\subsection{Interview Design}

This subsection describes our interview design and the goals of each question, which were informed by related cybersecurity research and refined by a pilot study. 

\subsubsection{Interview Questions} 
Each question was created to directly address one or more aspects of our \glspl{RQ}. Questions were organized into five thematic groups: background, training, military culture, mission impact, and command influence. We deviated from this order in the interviews to avoid leading questions by introducing topics early. Questions were informed by prior research: we used findings from \cite{raymaker2025sea} to develop more pointed questions about training (Questions 5-8) and a small subset of similar questions (Questions 3, 18, 19) for comparison, and \cite{maxam2024interview} guided scoping questions for military participants. The full, ordered sets of interview questions and scenarios are provided in Appendices~\ref{app:survey_qs} and~\ref{app:interview_qs}, respectively.

\paragraph{Background and Understanding}
These questions align with aspects of both \cref{rq:1,rq:2}. They seek to establish a baseline understanding of the participant's career and relevant cybersecurity experiences to address \cref{rq:1}. For \cref{rq:2}, they gauge the ways the participant identifies cyberattacks and gather their initial understanding of cybersecurity effects before later questions.

\paragraph{Training}
These questions primarily address \cref{rq:1} by identifying the participant's formal cybersecurity training. They also investigate mission influences and compare cyber training to physical threats, a robust training program. To answer \cref{rq:2}, they gauge the participant's preparedness in \gls{gps}-denied environments.

\paragraph{Military Culture} %
These questions investigate the effects of military culture and norms on cybersecurity perceptions. They address \cref{rq:1}, investigating position-specific responsibility, perceived advantages or disadvantages of military life, rank influence and indicators of expertise, as well as \gls{IT} and cybersecurity separation at units.

\paragraph{Mission Context} %
These questions answer both \cref{rq:1,rq:2}. For \cref{rq:1}, they explore the impact of missions, afloat vs. ashore postures, and military vs. merchant environments on cybersecurity. For \cref{rq:2}, they ask the participant to identify the cyberattack vulnerabilities and consequences they perceive as most consequential. 

\paragraph{Command Influence} %
These questions address \cref{rq:1}. They ask about specific command behaviors that relate to cybersecurity and to present their view of cyber responsibility in the organization.

\begin{figure}
    \centering
    \includegraphics[width=0.4\textwidth]{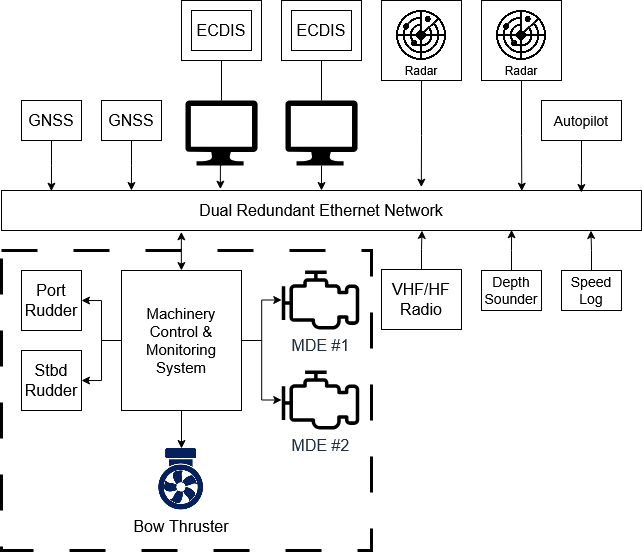}
    \caption{Example ship handout for scenario-based questions.\protect\footnotemark}
    \label{fig:example_ship}
    \Description[Ship Bridge Layout]{A figure showing the electronic navigation equipment layout on a ship's bridge. It shows two ECDIS consoles connected with two GNSS receivers, two RADARs, and autopilot via a dual ethernet network. The mechanical equipment includes two rudders, two MDEs, and a bow thruster. Sensors include radio, a depth sounder, and speed log.}
\end{figure}

\subsubsection{Scenario-Based Questions}
To examine how mariners reason about cyber risk in operational contexts and address \cref{rq:2}, we presented participants with three plausible scenarios: (1) a \gls{gps} spoofing incident while underway in clear weather, (2) a loss of propulsion control in a high-traffic area, and (3) a total failure of the ship’s \gls{ENAV} system while transiting a congested channel at night. These scenarios were selected to address the three overarching dependencies of ship safety, as described in authoritative nautical doctrine: positional certainty, situational awareness, and vessel control, respectively~\cite{cutler2004dutton, bowditch1906american}.

\footnotetext{ECDIS: Electronic Chart Display and Information System; GNSS: Global Navigation Satellite System; VHF: Very High Frequency radio; MDE: Main Diesel Engine}

For each scenario, participants were asked two core questions: what immediate actions they would take, and what indicators they would assess to diagnose the anomaly. These scenarios targeted distinct but critical shipboard systems (e.g., navigation, propulsion) that can be affected during a cyberattack.

To anchor responses in a shared operational context while avoiding sensitive disclosures, we provided participants with a reference ship configuration, shown in Figure~\ref{fig:example_ship} and Appendix~\ref{example_ship}. This notional vessel included realistic control systems, communications equipment, and navigation subsystems common to both civilian and military platforms. We chose these specific systems and redundancies to meet the requirements of international conventions and industry best practices~\cite{IMO_SOLAS_V_R19, IMO_MSC496_105, DNV_RU_SHIP_Pt6Ch2Sec7}. The example was not intended to be exhaustive, but offered a technical baseline that enabled participants to interpret scenario details, articulate decision-making processes, and reflect on how cyber disruptions might be handled.

\subsubsection{Pilot Study}
To refine the interview protocol, we conducted a mock interview with a researcher possessing substantial maritime domain knowledge, followed by a pilot study of three mariners. 
Participants consistently provided clear and substantive responses, and the protocol required only minor refinements: one of 27 questions was reworded for neutrality, and the question order was adjusted to better group topics and maintain a natural flow of questioning. No questions were added, removed, or substantively altered, and the responses to the reworded question did not differ in character before and after the change. Given the minor nature of these revisions, we include the pilot interviews in the final analysis.

\subsection{Recruitment, Surveys, and Interviews}
Recruiting active-duty military personnel presents well-known challenges, including a relatively small eligible population and restrictions on participation. We leveraged professional connections, snowball sampling, and LinkedIn outreach to recruit participants. We recruited 10 mariners through professional connections, including seven peers and three supervisors; no subordinates were recruited to avoid perceived pressure to participate. Snowball sampling allowed participants to recommend colleagues with relevant experience and gathered eight participants. LinkedIn outreach broadened recruitment beyond the authors' immediate networks with two mariners. The skew of recruitment sources reflects the difficulty of recruiting from military populations. All participants were uncompensated volunteers, as approved by the \gls{IRB}, to avoid a conflict of interest when leveraging professional connections.

To assess participants' eligibility for this study, we administered a survey through Qualtrics to collect background information and experience. We found that dedicated cyber officers cannot talk about their jobs in unclassified settings and chose to recruit general military mariners (service members who served aboard \gls{us} military vessels as an operator or as a technical expert maintaining ship-specific systems). Of note, some experts were not assigned to ships but worked onboard to manage specific systems. We included these participants as their role is representative of non-active duty expertise. Of the 30 mariners who completed the survey, 26 were invited to interview. Only 20 mariners completed an interview.

The interviews were conducted over Zoom, between August 2025 and January 2026, and lasted about one hour. Interviews were recorded and automatically transcribed with the participants' consent. All recordings were deleted after transcription to preserve anonymity, in accordance with our approved \gls{IRB} procedures.

\subsection{Participant Demographics}
Our study included 20 military mariners from the \gls{uscg}, \gls{usn}, and \gls{usmc}. Participants comprised 17 officers, one enlisted mariner, and two federal civilian employees who worked directly on military ships. 15 participants were affiliated with the \gls{uscg}, four with the \gls{usn}, and one with the \gls{usmc}. The \gls{usmc} member was embedded on a \gls{usn} ship. As such, we categorize them with the \gls{usn}. A total of 16 of the mariners held ship-driving roles as \glspl{ood}. We included the perspectives of both commissioned and enlisted personnel; however, this distribution demonstrates the operational focus of the study on decision-making and oversight. The skew of \gls{usn} to \gls{uscg} personnel reflects cultural barriers to recruitment across different services.

The participants' backgrounds were distributed across all major military accession sources, including federal service academies, civilian maritime academies and institutions, and enlisted pathways~\cite{usn_paths, uscg_paths}. Several progressed through the service, moving from enlisted to warrant to commissioned ranks, demonstrating a breadth of knowledge. The average age range was 30-34 years old, slightly higher than the average of \gls{us} service members (28 years old)~\cite{DoD2023Demographics}. Three participants were women, totaling 15\% of our study. Specific demographics of shipboard personnel are not available, but this makeup is representative of total service demographics, where women make up 20\% of the \gls{usn} and 16\% of the \gls{uscg}~\cite{DoD2023Demographics}. The participants had an average sea time of 4.7 years. Table~\ref{tab:demographics} shows the background of each participant. We did not identify participants' commissioning sources, past assignments, or genders to preserve anonymity, though this data was collected as part of our analysis.

\begin{table}[!t]
    \centering
    \caption{Participant demographics and backgrounds.\protect\footnotemark}
    \setlength{\tabcolsep}{4pt}
    \renewcommand{\arraystretch}{1.15}
    \small
    \begin{tabular}{@{}llccp{2.6cm}@{}}
        \toprule
        \textbf{ID} & \textbf{Branch} & \textbf{Rank} & \textbf{Yrs} & \textbf{Assignments Held} \\
        \midrule
    
        P1 & USCG & O2 & 2  & DWO, AOPS \\
        P2 & USCG & O3 & 4  & DWO, CSO, AOPS, OPS \\
        P3 & USCG & O2 & 3  & DWO, CSO, XO \\
        P4 & USCG & O2 & 3  & DWO, ANAV, XO \\
        P5 & USCG & O4 & 7  & DWO, XO, CO \\
        P6 & USN  & GS & 4  & Civilian SME \\
        P7 & USCG & O3 & 5  & DWO, AOPS, XO, CO \\
        P8 & USCG & O3 & 5  & DWO, CSO, XO \\
        P9 & USN  & E5 & 4  & Reactor operator \\
        P10 & USMC & O3 & 1  & USMC \\
        P11 & USN  & O2 & 1  & AEO \\
        P12 & USCG & O2 & 2  & EOIT, AEO, DCA \\
        P13 & USCG & O2 & 12 & DWO, ACSO \\
        P14 & USCG & O3 & 8  & DWO, XO, CO \\
        P15 & USCG & O2 & 3  & DWO, EOIT, XO \\
        P16 & USN  & O3 & 9  & OPS, WEPS \\
        P17 & USCG & O4 & 7  & DWO, CSO, CO \\
        P18 & USCG & O5 & 4  & DWO, EOIT, AEO, EO \\
        P19 & USCG & O3 & 5  & DWO, XO, CO \\
        P20 & USCG & GS & 0  & Civilian SME \\
        \bottomrule
    \end{tabular}
    \label{tab:demographics}
\end{table}

\footnotetext{Rank Acronyms: Officer (O), Enlisted (E), Civilian Federal Employee (GS). Assignment acronyms: DWO (Deck Watch Officer), EOIT (Engineering Officer in Training), EO (Engineering Officer), XO (Executive Officer), CO (Commanding Officer), CSO (Command Security Officer), OPS (Operations Officer), WEPS (Weapons Officer), SME (Subject Matter Expert), A (Assistant).}

\subsection{Data Collection and Analysis}
Military service members have access to sensitive information and can be targets, so care was taken to anonymize subjects and screen information. After interviews were audio-recorded and transcribed, we removed all identifying information from transcripts and an alphanumeric ID was assigned.

To ensure no classified or sensitive information was collected, participants were reminded of the unclassified nature of this research at the beginning of each interview. We followed the practices of~\cite{maxam2024interview}, reviewing each transcript before analysis occurred. Transcripts were securely stored and only accessible to approved researchers. These research procedures were approved by our \gls{IRB}.

We first used an iterative, inductive codebook methodology for thematic analysis with a multi-coder approach~\cite{braun2006using,srivastava2009practical, sahin2023investigating, wong2024comparing}. Then, we applied a reflexive process to develop and refine higher-level themes~\cite{braun2019reflecting}. Two researchers independently coded transcripts using a shared codebook that evolved through successive rounds of coding and reconciliation. As nuanced codes emerged, both coders reviewed all prior occurrences before deciding to combine or separate them and updated each occurrence accordingly. This process was repeated until the final codebook was established, which is available at an \gls{OSF} repository~\cite{osf_repo}. Our analysis produced 383 unique codes across 27 questions, including three scenarios. Intercoder reliability was calculated using Krippendorff's $\alpha$ to account for nominal, mutually inclusive data based on the application of the final codebook across all interviews. $\alpha = 0.883$, demonstrating consistent codebook application~\cite{hayes2007answering}. Because the codebook was refined and old codes updated through reconciliation at each interview and Krippendorff's $\alpha$ demonstrated excellent reliability, a final pass across all transcripts was not conducted. 

Higher-level themes were developed and refined through a reflexive process informed by Braun and Clarke's approach to thematic analysis~\cite{braun2019reflecting}. After the codebook was finalized, researchers grouped codes into initial themes and reviewed each candidate against the whole dataset, iteratively merging, splitting, and discarding candidates. When two candidates did not have distinct central concepts, they were restructured as sub-themes of a shared higher-level theme. This yielded five sub-themes under two overarching themes, corresponding to~\cref{rq:1,rq:2}.

We assessed code saturation to determine when a sufficient number of interviews had been conducted by following accepted standards in qualitative research~\cite{guest2020simple}. We tracked the emergence and the proportion of new codes per interview. Each code was assigned to one of a set of related code groups maintained during coding. No new groups emerged after the 18th interview, but we conducted two additional interviews. The proportion of new codes first fell below 10\% at the 14th interview. This percentage fluctuated slightly but remained under 13\% for the last six interviews, averaging 10.2\%. These new codes represented nuanced perspectives of existing groups. A plot of new codes over time and our identified saturation point is provided in Appendix~\ref{app:figures}. This trend signaled a comprehensive representation of ideas and confidence in our findings.

\subsection{Limitations}
\label{sub:limitations}
Our primary limitations are like many qualitative interview studies. The sample size of 20 participants limits claims about prevalence or generalization, but comparable sample sizes are common in interview studies~\cite{maxam2024interview, raymaker2025sea}. Our participant demographics span multiple services, ranks, operational roles, and commissioning pathways. All of these increase the range of perspectives captured, and we used saturation-based stopping criteria to support completeness.

Our findings are based on self-report rather than the observation of participants during live casualties or exercises. Additionally, participant responses to scenarios may reflect hypothetical or expected reasoning rather than in-the-moment behavior under operational constraints. Responses may also be subject to recall bias, as describing past incidents or perceived vulnerabilities can be incomplete. To mitigate this, we followed well-accepted practices for interview creation, including refining questions after a mock interview and pilot study of three participants.

Because this study was conducted entirely at the unclassified level, participants could not discuss exact shipboard response protocols, system configurations, vulnerabilities, or classified threat intelligence. Therefore, our findings reflect operators' unclassified mental models and practices, and may omit procedures or capabilities used in classified environments. This concern cannot be mitigated given classification boundaries. However, we argue that addressing the lack of systematic research on military cybersecurity requires studying unclassified procedures, where access is feasible and insights can be reused by broader security communities.

Additionally, our sample was limited to \gls{us} military personnel, so our findings may not generalize to the cybersecurity practices, doctrine, or culture of other nations' militaries.

Finally, recruitment relied heavily on professional networks and snowball sampling, which may bias the sample toward more engaged participants. To reduce this risk, we screened participants for eligibility and role diversity to ensure participants represented both services and a variety of communities.

\section{Shaping Cybersecurity Risk Perception} %
\label{ch:4_prep_and_practices}

This section addresses \cref{rq:1}: \textit{What factors shape \gls{us} military mariners' mental models of cybersecurity risk?} We first examine how learning factors, including formal training and other experiences, shape cyber knowledge. Then, we review how military organizational structures influence responsibility and accountability and analyze how military mariners use situational factors to shape cyber risk. These factors informed our first overarching theme: because cybersecurity is organizationally abstract (deferred in priority, ambiguous in responsibility, and judged by context rather than doctrine), military mariners shape their models of cyber risk from informal experiences and situational factors rather than standard instruction.

\subsection{Fragmented Learning Pathways}
\label{sub:learning_pathways}
Here, we present the formal training and informal experiences that influence participants' cyber understandings. We draw insights from questions about training and background.
\subsubsection{Training:}
\label{subsub:training}
Nearly all, 17 of 20, mariners reported receiving the same annual online cybersecurity training as part of mandated requirements, regardless of platform. Known as the \gls{dod} Cyber Awareness Challenge, this training is a baseline annual requirement for all \gls{us} military personnel~\cite{dod_cyber_challenge}. 13 of 20 described this training as largely irrelevant to shipboard duties, often characterizing it as a check-the-box requirement that does not reflect shipboard environments. One participant attributed this mismatch to the constraints of security classification: \enquote{\textit{Most of the cyber threat information that we have for afloat assets is at the [top secret] level. And it's unfortunate that, generally, the [large cutter] commanding officers are the only ones to have that clearance}} (P18). This classification barrier creates an obstacle to effectively describing cybersecurity threats, disconnecting formal training and operational reality for mariners without sufficient clearance.

Despite its limited shipboard relevance, 6 of 20 participants viewed the training as useful for establishing basic computer-use habits, a crucial skill for junior personnel entering the service. 7 of 20 independently identified risks associated with connecting personal devices (e.g., phones or USB drives) to government systems, citing this as one of the few concrete lessons reinforced by formal training. One participant described a clear behavioral shift over time: \enquote{\textit{I haven't seen anyone plug an iPhone into a Coast Guard network computer in about seven or eight years. It used to happen every reserve weekend. Every. Every single one}} (P8). This account suggests that formal training can still shape safe baseline practices.

Only 6 of 20 participants reported cyber training that was tailored to their unit's mission, including region-specific threats or mission-specific risks. The remaining 14 mariners described no mission-specific training at all. When contrasting cybersecurity training with physical threat response programs, such as military \gls{DC} or \gls{ATFP} programs, participants consistently highlighted a disparity. 17 of 20 characterized their cybersecurity training as absent, low-quality, or low-priority. Conversely, 13 of 20 participants described physical threat training as robust and taken seriously. Physical threat training was repeatedly cited as hands-on and effective with clearly defined threats and expectations: \enquote{\textit{[It] was daily, sometimes multiple times a day... And [it] was taken very seriously. I personally had fires and flooding, and I can say that the training works. I'm still here, the boat didn't sink, and everybody did their job}} (P9). This disparity suggests that cybersecurity is not treated as an operational requirement with the same focus as rigorous physical threat training.

Participants also identified a consistent gap in what shipboard systems training addressed. Nearly all participants (18 of 20) stated that their cybersecurity training did not distinguish between traditional \gls{infotech} systems and \gls{OT}, like propulsion and navigation equipment. Only 2 of 20 (both members of the Navy) identified training with a clear distinction between systems. Of them, one received it through direct instruction and the other described it as the inclusion of network-specific training into mechanically inclined careers, like mechanics and electricians. 

These responses indicate a persistent misalignment between cybersecurity training and shipboard environments. Operations prioritize well-defined physical threats with robust training in a way cybersecurity has not yet achieved. However, a minority view suggests that baseline cyber training still reinforces some habits.

\subsubsection{Experiential Factors:}
\label{subsub:informal_influences}
Informal learning emerged as a dominant source of understanding. Only 2 of 20 participants cited formal online cybersecurity training as a major influence on their cybersecurity understanding, despite its uniform presence across the study. Instead, participants described a wide range of informal influences that were often unstructured and opportunistic. We identified 24 unique codes related to cybersecurity influences, indicating considerable heterogeneity. Only four influences appeared more than twice: using electronic devices (5 of 20), college courses (4 of 20), working with the Navy (3 of 20), and news media (3 of 20). At a higher level, these influences clustered into two mutually inclusive themes: 14 of 20 participants pointed to experiences had within the military, while 8 of 20 described factors originating outside military training. These patterns indicate that mariners' cybersecurity perceptions are shaped more by experience-driven exposure than formal training; however, many learn from experiences in the military that they would otherwise not receive.

A small number of participants described particularly influential assignments that contributed to deeper cybersecurity learning. For example, one participant leveraged their position as \gls{CO} to initiate shipboard cybersecurity discussions about relevant cyber vulnerabilities with watch standers, effectively prioritizing learning at the unit level. Five participants described relevant cybersecurity backgrounds, such as federally sponsored intelligence programs or formal degrees, that provided extremely relevant, ship-specific knowledge. However, these backgrounds were viewed as exceptions rather than the norm.

These findings show how cybersecurity knowledge is developed in the absence of operationally relevant training. Still, Section~\ref{subsub:training} suggests that standard training fosters valuable baseline habits.

\begin{tcolorbox}
\textbf{Military mariners build cybersecurity understanding through fragmented, experience-driven pathways, suggesting gaps in standardized instruction.}
\end{tcolorbox}

\subsection{Organizational Responsibility}
\label{sub:inconsistent_resp}
Next, we examine how military organizational structures and roles shape cybersecurity responsibility. We draw insights from questions about chain of command, rank, role-based responsibility, and other factors that shape expertise.

\subsubsection{Chain of Command:}
\label{subsub:chain_of_command}
Participants framed cybersecurity as a multi-level organizational issue. When asked about command behaviors and cybersecurity responsibility, participants separated the military's organizational hierarchy into three levels, representative of their rigid chain of command: top-level leadership (headquarters), individual units (a \gls{CO}), and the individual member.

At the top, participants shared beliefs that headquarters priorities and policy drive the organizational focus. 7 of 20 participants found the policy-driven nature of the military to complement cybersecurity, as strict policies, additional resources, and remedial training make compliance the norm. Participant P4 shared a common compliance-rooted mentality: \enquote{\textit{The fact we are in the military and are absolutely required to follow policy where possible. That makes it a lot easier [to care].}}

At the command level, participants described many positive and some detrimental behaviors. Nearly every mariner (18 of 20) described commands encouraging compliance with key policies and good computer habits. Poor command-level practices were rare but still arose. 4 of 20 reported an absent or reactionary cybersecurity culture onboard, such as prioritizing operations over security. One participant noted a perceived positive shift in commands' mindsets toward cybersecurity assessments: \enquote{\textit{It used to be very resistant culture. \enquote{Why is cyber coming on board? To ruin my life?} ... I think now people are very open to it because they realize it's better to find out now, rather than to find out when you get potentially hit with a zero-day attack}} (P18). This perspective suggests a growing acceptance of proactive cyber assessments.

The individual level was deemed most important: most mariners (15 of 20) identified cybersecurity as primarily an individual or shared responsibility. These mariners emphasized the importance of the human factor, describing how strict organizational controls cannot fix unsafe habits. One participant emphasized these personal actions: \enquote{\textit{[Cybersecurity] begins with us, every individual. All the cyber awareness stuff, \enquote{don't click on those strange emails,} \enquote{don't share your passwords,} \enquote{don't leave your [ID] in.}}} (P14) 

\subsubsection{Rank:}
\label{subsub:rank}
Military organizations operate under rigid rank hierarchies, but participants did not consistently view rank as a proxy for cybersecurity expertise. While 3 of 20 participants suggested that concerns raised by higher-ranking personnel would be addressed more quickly and 4 characterized cybersecurity as primarily a senior-level responsibility, only one participant positively correlated rank with cyber expertise. In contrast, 10 of 20 participants did not think rank influenced their assessment of cybersecurity credibility, and 4 argued that higher rank could negatively affect perceived expertise. One participant observed that rank disparities may deprioritize cybersecurity altogether: \enquote{\textit{The department head that deals with all things cyber [is a lower rank], so I have a feeling cyber might get punted to the side a fair amount}} (P12). These accounts suggest that although rank strongly shapes organizational decision-making, it is rarely viewed as cybersecurity competence.

\subsubsection{Expertise Indicators:}
Consequently, we observed that participants rely on varied indicators when gauging cyber expertise. They provided 17 unique signs of expertise, the most common of which included formal education (6 codes), job experience (6 codes), and cybersecurity certifications (5 codes). Responses were evenly divided among formal civilian indicators, military indicators, technical experience, and demonstrated skill. This variability of responses suggests a lack of consensus when gauging expertise among shipboard personnel. \enquote{\textit{I would say, in general, ... it's just experience, certifications, qualifications, and so forth. But those are not things that I think you would typically find aboard any vessel, whether it be a Coast Guard cutter or a merchant vessel}} (P1). The ambiguity of these indicators suggests that military mariners cannot clearly gauge expertise because organizational roles are unclear.

\subsubsection{Role-Based Responsibility:}
\label{subsub:positional_resp}
We asked participants how shipboard billets\footnote{Billet is a military term for a role that requires specific duties and skills.} shape cybersecurity responsibilities. Only two codes occurred more than twice: their billet was not responsible for cybersecurity (8 of 20) or that other specific billets were responsible for cybersecurity (5 of 20) (e.g. \gls*{OPS}, \gls*{CSO}, \gls*{NAV}). Of note, these billets have primary responsibilities that are not cybersecurity-focused. Participant P7 emphasized a lack of cyber responsibility as a \gls{CO}: \enquote{\textit{My assignment as CO means that I need to consider all security related issues to the cutter, which includes cyber, but there's nothing in the Coast Guard that requires me to take those things into consideration as much as ATFP, DC...}} Another discussed the creation of a dedicated cyber officer role aboard naval vessels: \enquote{\textit{The Navy has been working on this for years. It took them a long time to do it, but some of their ships have a junior officer [who] is the \gls{CYBO}... [They] report to the commanding officer every day with a cyber report.}} (P18) 

Participants also demonstrated limited differentiation between traditional \gls{infotech} support and cybersecurity roles. 9 of 20 participants reported a poor distinction between general \gls{infotech} assistance and cybersecurity support onboard vessels. Another 2 of 20 described that only certain technical experts were assigned onboard ships and, thus, \gls{infotech} professionals often became cyber liaisons. Further, one participant's insight revealed blurred distinctions at the organizational level, where \gls{infotech} support staff were governed by cybersecurity structure: \enquote{\textit{The people you called to get \gls{infotech} support... were actually governed under our cyber people. I think we've realized that was a bad idea, ... and we're moving them back out... because they are \gls{infotech} support. They're not cybersecurity folks.}} (P18) 

This implies that when mariners were talking with \gls{infotech} support, they were unknowingly going through cybersecurity channels. While corrective action was reported, the conflation suggests that cyber responsibilities may be poorly delineated at a larger level.

Together, these accounts illustrate ambiguous responsibility across military structures, which may complicate accountability and response. They suggest that dedicated billets (e.g. \gls{CYBO}) can make cybersecurity actionable in ways that informal ones cannot.

\begin{tcolorbox}
\textbf{Military mariners suggest that cyber responsibility is inconsistently applied in strict organizational hierarchies, potentially complicating accountability.}
\end{tcolorbox}

\subsection{Context-Dependent Risk Perception}
\label{sub:context_dependent}
We next examine how military mariners use situational factors to adapt their cyber risk profiles to fit operational realities. Specifically, we compare military versus civilian operations, afloat versus ashore environments, and different mission types.

\subsubsection{Military vs. Civilian:}
\label{subsub:military_civ}
Participants frequently identified differences in military assets and structure that contributed to improved cyber resilience. When comparing the security of military and merchant vessels, 13 of 20 participants identified structural advantages like better resources, protected and redundant systems, and dedicated cyber response teams that contribute to improved security. One participant gave firsthand insight of securing both military and merchant ships: \enquote{\textit{Compare a national security cutter, that was designed with cyber in mind from day one, to [a civilian ship]... I did both assessments, and it's a night and day difference}} (P18).

Despite the widespread acknowledgment of resource advantages, many participants did not recognize the extensive hardening the systems undergo. When asked for their opinion on whether military vessels are more resilient than merchant ones, 14 of 20 mariners labeled the military as more secure, but only 9 of these grounded their belief in clear reasons, like increased funding or dedicated cyber response teams. 5 participants could not provide specific reasons, with some choosing to trust the government to oversee cybersecurity. When asked about this knowledge gap, a mariner with experience securing shipboard systems highlighted the challenge of frequent crew turnover: \enquote{\textit{They had to go through all of the requirements and all of the testing with us. So those crews that take over ships [from the shipyard] have a pretty good idea about the effort, but once those crews change out, once the new commanding officers change out, it's a whole lot harder}} (P18). The entire crew of a military ship typically cycles out every two to three years~\cite{usn_tours, uscg_tours}. This suggests that even when platforms are tested for resilience, rapid personnel turnover can erode the unit-level understanding.

Separately, we asked mariners what they perceived as the most significant consequences of cyberattacks against both military and merchant vessels. Participants identified severe consequences of cyberattacks against both types of vessels, though they differed by context. For merchant vessels, participants most frequently emphasized economic concerns (10 of 20), loss of the vessel (9 of 20), and loss of life (7 of 20). When considering military vessels, participants similarly cited loss of life (8 of 20) but also highlighted unique consequences to military operations. These included hijacking weapons systems (7 of 20), geopolitical escalation (7 of 20), degraded operational availability (6 of 20), and loss of data confidentiality (4 of 20). Participant P3 noted that many consequences, such as entering another nation's territorial seas without clearance, do not translate to merchant vessels: \enquote{\textit{For merchant ships, it's usually a rite of passage, whereas military ships are a show of force, to some extent, and that might not be well received by whatever country we [enter] their territorial seas.}} The most cited and consequential attacks against military vessels involved losses of system integrity. False positioning information could lead a military vessel to violate the territorial sovereignty of another country. \gls{gps} spoofing is a widely speculated cause for the Iranian seizure of two \gls{usn} boats in 2016~\cite{usn2016spoofing}. Taking over a weapon system through direct control or falsified targeting commands was presented as another integrity-based attack: \enquote{\textit{You're going to have vessels that are in charge of others. If you compromise the integrity, then what are we doing here? You could cause someone to shoot at something that shouldn't be shot at.... You could cripple a vessel or you could cause World War III if you wanted to}} (P1).

We found these higher consequences of cyberattacks against military vessels to contribute to participants' cyber understandings.

\subsubsection{Environment:}
\label{subsub:afloat_ashore}
To analyze the effects of ship-specific environments and constraints, we asked participants to compare the perceived cyber readiness of afloat and ashore units. Interestingly, no participants identified ships as more secure. Conversely, 10 of 20 participants found ashore units to be more secure, citing additional resources on land or ship-specific consequences. One participant shared an example of vulnerability-inducing behavior on a vessel that lacks consequences on land: \enquote{\textit{I remember the \gls{CO} said, \enquote{you will not make Wi-Fi networks using your video games on the submarine again.} Because you could find the Wi-Fi networks, right? They had reasons, it degrades other signals, communications, and stuff like that, but the big picture was it left us vulnerable}} (P9). A competing mentality, shared by 5 of 20 participants, found no correlation. They attributed differences to other factors like available systems and missions, where operational units were seen as having higher readiness than supporting units. Overall, these comparisons suggest that afloat cyber readiness is perceived as constrained by the operational realities of ships, where limited resources and isolated operating environments introduce unique challenges.

\subsubsection{Mission:}
\label{subsub:mission}
Cyber readiness almost always changed with the mission at hand; only one participant noted no change. Higher levels of readiness were primarily associated with changes in \gls{OPSEC} or adversary sophistication (9 of 20 participants) and \gls{EMCON} (4 of 20). Participants described near-peer adversaries as the most malicious, followed by drug-smuggling operations and violators of international law. Operating near these adversaries was perceived to require heightened postures, while lower-threat missions (e.g., aids-to-navigation or ice-breaking) permitted reduced security practices. \gls{OPSEC} is a structured process for protecting sensitive information by identifying critical information and actions that an adversary could collect and exploit, then applying countermeasures to reduce that risk. \gls{EMCON} similarly provides concrete direction by selectively limiting electromagnetic emissions to reduce detection by adversarial sensors and avoid interference, turning abstract security goals into specific actions. Together, \gls{OPSEC} and \gls{EMCON} give operators tangible, procedural steps to follow depending on the mission at hand. While not strictly cybersecurity postures, both contribute to the readiness of systems and recurred throughout interviews.

While cyber readiness changed to suit the mission, operational availability and cybersecurity were frequently framed as competing priorities. When participants were asked how military life affects cybersecurity, 12 of 20 shared ways operational demands and mission priorities contradicted cybersecurity. Participant P6 noted, \enquote{\textit{In some cases it was even seen as an obstruction to your duties. Sometimes we'd have what was needed to get a ship running, but because of cyber policies, we can't [install] the equipment. The reaction was always, \enquote{Can we turn around, close our eyes, and you do it anyway?} And I'd have to say no.}} This illustrates how cybersecurity can be treated as a challenge to work around. 

\begin{tcolorbox}
\textbf{Perceived cyber posture is contextual: military ships seem more secure than civilian, afloat units less so than ashore, and readiness shifts with missions.}
\end{tcolorbox}

\section{Military Mariners' Operational Response}
\label{ch:5_mental_models}

\begin{figure*}
    \centering
    \includegraphics[width=\textwidth]{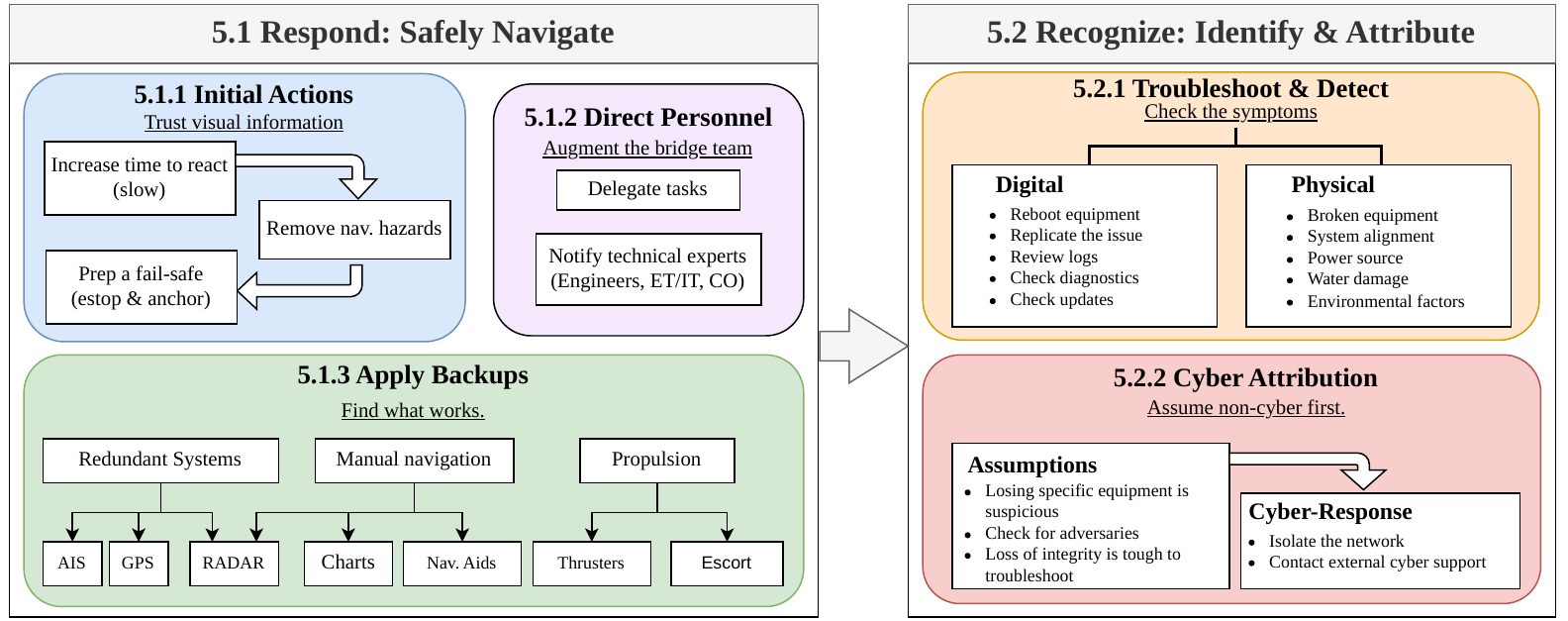}
    \caption{Safety-oriented incident-response model of military mariners.}
    \label{fig:process}
    \Description[Response Model Figure]{A figure illustrating the response model of military mariners. The left half, part 1, covers safety of navigation. Its sections include initial actions, directing personnel, and applying backups. Part 2, on the right, covers identification and response. It includes troubleshooting and cyber attribution.}
\end{figure*}

In this section, we answer \cref{rq:2}: \textit{How do \gls{us} military mariners recognize and respond to cybersecurity threats under organizational constraints?} We analyze how mariners responded to cybersecurity threats in three operational scenarios and present a safety-oriented incident-response model. We present this model in the order participants applied it: starting with navigational actions then moving to cyber-specific identification and attribution. We end by comparing mariners' understood vulnerabilities against their detection methods to assess alignment. These factors informed our next theme: military mariners make cybersecurity actionable by recognizing operational impacts and responding with practiced seamanship, creating resilience but possibly delaying cyber attribution.

\label{sub:op_response}

Here, we analyze how operators respond to and mitigate cybersecurity threats in operational environments with insights from participant responses to the three scenarios, each illustrating a potential cyberattack on a critical shipboard system: \gls{gps} spoofing, a loss of propulsion control, and a sudden outage of \gls{ENAV} systems. Participants consistently followed a structured decision-making process focused on safety and operational continuity. This process model reflects longstanding maritime doctrine, which prioritizes immediate vessel safety, redundancy through independent navigation methods, and coordinated bridge team actions before complete system diagnosis~\cite{cutler2004dutton,bowditch1906american}. We aggregated each participant's responses and connected them to these established principles of navigation to model how military mariners approach cyber incidents. This model is illustrated in Figure~\ref{fig:process} and is separated in two stages: respond and then recognize. The following sections detail each stage in turn. As noted in Section~\ref{sub:limitations}, responses here may reflect hypothetical or expected reasoning rather than their actual behavior. (Unless otherwise noted, counts in this section are of the 16 participants who chose to answer scenario-based questions.)

\subsection{Respond: Safely Navigate}
\label{subsub:safe_nav}
The first stage of the process model prioritizes immediate response to maintain positive control of the vessel before deeper system understanding. These behaviors are unique to maritime and aviation environments, where vessel movement can endanger lives and create urgency. Nautical doctrine teaches mariners to navigate first to reduce risk exposure before diagnosis~\cite{cutler2004dutton, bowditch1906american, adams2010shipboard}.

\subsubsection{Initial Actions:}
These are the \gls{ood}'s first steps to mitigate risk. They are immediate, almost instinctive measures taken before any amplifying information is known. All but one participant described actions that could be categorized as ensuring safe navigation, emphasizing it as the participants' first priority. This was often achieved by slowing down to allow more time to react and through removing navigational hazards, by navigating to anchorage areas or open water. In total, 9 of 16 participants stated they would notify nearby vessels to stay clear because their maneuverability might be impaired, demonstrating situational awareness rather than problem-fixation. Most participants (10 of 16) also discussed preparing the anchor as a fail-safe for a complete loss of control. These most common actions directly align with the \enquote{aviate} principle of ship driving, focused on maintaining control.

\subsubsection{Direct Personnel:}
Next, 10 of 16 mariners sought to augment the navigation team with extra personnel, either technical experts (e.g. engineers or the \gls{CO}) or navigation team members. This delegation aligns with principles of bridge resource management, which emphasize the need for fine-grained separation of duties in dynamic or challenging operations. The separation is meant to allow each person to carry out a specific task without distractions, mitigating potential task-fixation~\cite{adams2010shipboard}.

\subsubsection{Apply Backups:}
A core assumption of nautical doctrine is that no single navigation system is fully reliable~\cite{cutler2004dutton, bowditch1906american}. 12 of 16 participants explicitly discussed transitioning to redundant systems, emphasizing this assumption. Each identified a subsequent backup system to replace the affected one. In scenarios that affected navigation equipment, \gls{ENAV} was backed up by redundant \gls{gps} receivers, isolated \gls{ENAV} systems, and manual techniques. The ultimate backup, called \enquote{seaman's eye}, is a way of using experience and observations to navigate without any alternatives. 5 of 16 participants used seaman's eye during the loss of \gls{ENAV} scenario.

We also asked about participants' comfort when sailing in \gls{gps}-denied environments to better understand their perception of a widely perceived threat. Interestingly, 5 of 16 expressed outright discomfort in operating without \gls{gps}, viewing it as a major risk. Backups often relied on traditional navigational skills, which several perceived were negatively impacted by recent reliance on electronic navigation. One mariner described new navigation systems meant to address this issue: \enquote{\textit{We have systems on board cutters now that... have advanced navigation technologies built in to take pictures of the night sky... and be able to plot fixes and courses on electronic charting systems using celestial}} (P20).

Participants rarely investigated the affected system during their initial response. Most treated it as wholly inoperable, choosing not to troubleshoot until navigation or propulsion was at least partially restored: \enquote{\textit{We'd be focusing on getting to anchor, and going from there}} (P7). These behaviors align with established principles, where manual navigation is an expected contingency and more certain than immediate troubleshooting. Bowditch emphasizes that when discrepancies arise in navigation systems, mariners should not immediately diagnose the inconsistency but instead assume capabilities are degraded and navigate conservatively until confidence is restored~\cite{bowditch1906american}.

\begin{tcolorbox}
\textbf{Military mariners use a safety-oriented incident-response model that centers on preserving navigation, coordinating personnel, and using backups.}
\end{tcolorbox}

\subsection{Recognize: Identify \& Attribute}
\label{subsub:id_response}
The second stage of the model captures how military mariners transition from ship control to diagnosing and managing the underlying cause of an anomaly. This aligns with maritime doctrine, which prioritizes diagnosis only after immediate safety is addressed.

\subsubsection{Troubleshoot:}
The indicators and troubleshooting techniques participants identified can be divided into digital or physical actions, depending on the scenario. Troubleshooting was often described as being performed by the onboard experts and would be used to make an assessment about the cause and mission impacts. Participants identified valuable GPS receiver diagnostic techniques during the potential \gls{gps} spoofing incident: 6 of 16 described checking the number of satellites, signal strength, or the position estimate's dilution of precision. The remaining troubleshooting steps were less specific, though, indicating an uneven focus of knowledge. Most participants described routine, best-effort actions like looking for physical damage (15 of 16), rebooting the system (10 of 16), and checking the power (5 of 16). \enquote{\textit{They would probably try to power cycle it, it's usually our go-to option for all casualties on board}} (P3). When discrepancies arise between instruments, mariners are trained to investigate through cross-checking and comparison~\cite{cutler2004dutton, bowditch1906american}. These troubleshooting methods are standard seamanship practices applied to digital systems, rather than cyber-specific actions.

\subsubsection{Cyber Attribution:}
\label{sub:vuln_detect}
When speculating about the cause of each scenario, participants provided varied assumptions about cybersecurity. 2 of 16 participants stated they would assume non-cyber issues, while another 2 shared a perceived difficulty troubleshooting if system integrity had been compromised. \enquote{\textit{I'd assume it was anything but cyber at first. It'd probably take me quite some time to get to that conclusion}} (P2). Conversely, one participant stated they would assume a cyberattack during the \gls{ENAV} outage, noting that only losing critical navigation screens would raise suspicion. If cyberattacks were suspected, participants reported contacting and relying on external, off-ship cyber support. A participant with experience on cyber triage teams shared, \enquote{\textit{In my experience, something like that would result in a call to us, and someone would be on the next flight... to see what was going on}} (P6).

\paragraph{Misaligned Vulnerability and Detection Models} To better understand how military mariners recognized cyberattacks, we asked additional questions about understood cyberattack vulnerabilities and detection methods. By comparing these factors, we assessed the alignment of mariners' vulnerability-detection models. (Counts below are of the full sample, N=20, and are not mutually exclusive.)

We found the primary detection heuristic for potential cyberattacks to be operationally based: 18 of 20 participants described identifying attacks through obvious, observable failures. 11 of 20 would look for technical indicators (intrusion detection systems or high resource utilization), and only 6 emphasized formal cross-checking and sensor validation. Stealthy or covert attacks were identified as difficult to detect, as 4 of 20 participants cited limited diagnostic tools and uncertainty about what indicators would be visible to ship operators. One participant described this challenge to discern cyber threats from benign anomalies: \enquote{\textit{it's just keeping an eye out for something that doesn't look right}} (P7). These responses suggest that shipboard cyberattack detection is dominated by visible anomalies and operational effects.

\begin{figure}
    \centering
    \includegraphics[width=0.45\textwidth]{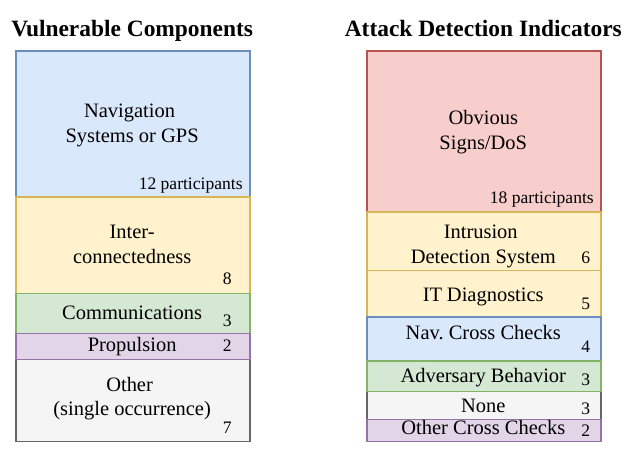}
    \caption{Cyber vulnerability and indicator frequencies.\protect\footnotemark}
    \label{fig:vuln_indicators}
    \Description[Vulnerable Component vs. Detection Indicators]{A figure comparing the frequency of vulnerable components, as mentioned by participants, against the frequency of attack detection indicators. The vulnerable components are navigation systems, including GPS (12 participants), interconnectedness (8), communications (3), propulsion (2), and other (7). Attack detection indicators are: obvious signs (18 participants), intrusion detection systems (6), IT diagnostics (5), navigation cross checks (4), adversary behavior (3), none (3), and other (2).}
\end{figure}

We asked what participants perceived as the most significant vulnerabilities aboard modern ships to understand how they prioritize cyber risk. They emphasized navigation-based attack surfaces, with 12 of 20 citing \gls{gps}, \gls{AIS}, and \gls{ENAV} equipment. System interdependence was also cited by 8 of 20 participants. Interestingly, 3 of 20 identified risks to propulsion or \gls{flss}, highlighting a rare but broader understanding. \Cref{fig:vuln_indicators} compares how often participants discussed these vulnerabilities to the attack indicators described above. Mariners' threat models emphasize navigation systems and interconnected devices, but their detection methods rely on obvious operational failures or \gls{infotech}-centric signals that are difficult to observe in shipboard environments. 

\footnotetext{Colors are used to correlate related vulnerabilities and detection indicators (e.g., navigation vulnerabilities are detected through navigation cross checks).}

\begin{tcolorbox}
\textbf{Mariners' perceived vulnerabilities and actual detection methods are misaligned, leaving detailed troubleshooting and cyber attribution to off-ship support.}
\end{tcolorbox}

\section{Concluding Discussion}

In this study, we examined how military service members understand and respond to cybersecurity threats. We focused on the experiences of personnel from Navy and \gls{cg} vessels, interviewing 20 mariners to identify recurring themes. We found service members to follow structured steps when responding to equipment anomalies, reflecting strong training and standardization. However, their threat models are misaligned with cyber risk: mariners focus on navigation disruptions, while integrity compromise poses a more significant risk in military operations. Their detection heuristics target overt failures instead of integrity compromise.

We identified unique cybersecurity consequences aboard military vessels and the existence of extensive platform hardening. Mariners demonstrated a baseline understanding of cybersecurity and adherence to strict policies; however, recurring training gaps remain. We synthesize \Cref{ch:4_prep_and_practices,ch:5_mental_models} to provide actionable recommendations for the broader international military community and civilian mariners, grounding them in our results. 

\subsection{Core Recommendations}
Here, we develop four core recommendations that address gaps in operator mental models and training. These recommendations apply broadly across the \gls{us} military, international services, and civilian maritime domains, with domain-specific implications being discussed in subsequent sections.

\subsubsection{Clarify Tangible Cyber Risk} 
The cyber vulnerabilities and attack consequences described by mariners in Section~\ref{sub:vuln_detect} varied without clear training and seemed to be formed by the many experiential factors discussed in Section~\ref{subsub:informal_influences}. This finding aligns with prior usable security research, which identified many users' reliance on prior experiences~\cite{wash2021knowledge} and a lack of adequate \gls{OT} training~\cite{evripidou2024understanding}.

Operators do not need to know sensitive vulnerabilities, but they would benefit from credible explanations of what manipulations are possible, the resulting consequences, and how benign failure differs from adversarial interference. Chen et al. found the expectation to mitigate attacks and understanding of consequences to be motivating factors when reporting institutional phishing emails~\cite{chen2024motivates}. Where feasible, exposing sanitized versions of cyber assessment processes and common threat vectors could help align military mariners' threat models of cybersecurity and increase its relevance.

\subsubsection{Establish Initial Cyber Actions}
Our safety-oriented incident-response model in Section~\ref{sub:op_response} shows that military mariners follow a consistent response for anomalous system behavior: stabilize the ship, gather people, then rely on redundancies before diagnosis. Section~\ref{subsub:safe_nav} describes how the first stage, \enquote{safety of navigation,} is rooted in nautical doctrine. It appears to be shaped by the effective physical threat response training described in Section~\ref{subsub:training}, where mariners are taught to decisively act against observable threats by mitigating the symptoms, like fire, before identifying causes. 

In Section~\ref{subsub:id_response}, we found that ships traditionally rely on external resources for cyber support. This affirms prior studies~\cite{raymaker2025sea, demchak2025cyber, rajaram2022guidelines} and nautical literature~\cite{cutler2004dutton, bowditch1906american}, much of which is tailored to merchant shipping. However, this approach is increasingly inadequate for military vessels, where adversaries are \textit{expected} to be intentionally malicious and subversive. Consider propulsion-related anomalies as an example: participants generally described diagnosing the physical systems before considering cyber attribution. If the underlying cause is a cyberattack, this delay in attribution could allow an attacker to retain control with severe operational consequences. 

Cybersecurity training should be integrated into these established safety-oriented response frameworks, particularly in military settings where ample resources are available. Establishing a set of clearly defined initial cyber actions can help operators respond decisively. These might include basic containment and input verification procedures (e.g., isolating subsystems when feasible), paired with clear reasons for when to take them. The emphasis should be on simple, repeatable actions that complement, rather than compete with, existing casualty responses to prepare entire crews for cyber threats when external support might not be immediately available.

\subsubsection{Normalize Cross-Checking}
Participant experiences pointed to integrity breaches, not outages, as a central military cyber risk in Section~\ref{subsub:military_civ}. While availability-based outages were frequently cited as disruptive, crews have well-developed practices for mitigating equipment casualties. The more dangerous failures occur when systems are available but untrustworthy~\cite{bhatti2017hostile}. If the operators cannot distinguish false data, they may never transition to redundant systems and instead act on compromised inputs, like falsified targeting commands or spoofed positions. Here, the strategic consequences can be more severe than a loss of operational availability.

This may be caused by mismatches between perceived consequences and attack-detection techniques mariners described in Section~\ref{sub:vuln_detect}. The most frequently cited detection heuristics were obvious, observable failures. Fewer participants mentioned the difficulty of identifying attacks. This idea mirrors outdated cybersecurity challenges that prompted the adoption of intrusion detection systems~\cite{goodall2004work, denning1987intrusion} and suggests a gap: mariners may overestimate their own detection ability, especially when systems appear functional. 

Establishing training and operator-friendly tools that emphasize cross-checks for verification across systems, not just navigation equipment, can help address this gap. Designing engineering systems to retain key analog components amid increasing automation would allow operators to intervene during casualties~\cite{danielsen2021still}. Rather than relying on breakdown, implementing these techniques would build proficiency and defend against integrity-based attacks. 

\subsubsection{Formalize Responsibility}
Rank was not viewed as a reliable indicator of expertise, and few shipboard positions were identified as being responsible for cybersecurity in~\Cref{subsub:rank,subsub:positional_resp}. Instead, credibility and authority were shaped by individual perceptions, leading to inconsistent views about what signals expertise.

With this culture, military services created a dedicated onboard cybersecurity role for some ships. This represents an important organizational step, but participants' reactions suggest its authority and purpose are not clear. One participant praised the creation of the role as a mechanism of clear ownership, while another critiqued mismatches between the position's rank and perceived responsibility. Most were oblivious to the billet entirely. Together, these perspectives emphasize the need for clearer delineation of cyber-specific functions in shipboard organizations and the alignment of roles to the reality of operational decision-making.

\subsection{Broader Military Applications}
These core recommendations are derived from the experiences of mariners within the \gls{usn} and \gls{uscg}. They are most directly suited for military maritime operations. However, the lack of prior research examining the experiences of service members in other \gls{us} or international military branches limits our ability to assess how these findings generalize across domains.

Many of the underlying challenges identified in this study, including increasing automation, reliance on networked platforms, and operation in adversarial environments, are not unique. Prior work on cyber operations and military cyber readiness suggests these challenges extend beyond maritime contexts, and cyber defense is an established, cross-domain military concern~\cite{libicki2009cyberdeterrence}. Further, the small body of international military maritime research supports the reliance on external support for cybersecurity and the need to operationalize cybersecurity~\cite{forson2022assessing}. We expect these recommendations to broadly apply across military branches, though their implementation may vary with domain-specific constraints.

\subsection{Civilian Maritime Applications}
Focusing on maritime services of the military afforded us a comparison with the civilian maritime sector. We compared findings from Raymaker et al. against insights from our participants to illustrate these differences in Table~\ref{tab:compare}~\cite{raymaker2025sea}, finding three main areas of distinction: threats discussed, perceived impacts, and asset mitigation in place. Navigational threats were disproportionately emphasized by both populations. We compare response content but lack a civilian baseline for response processes, since our scenario-based questions were designed specifically for military mariners.

Raymaker et al. includes some military-affiliated participants, but the number of active-duty mariners represented is low, and its questions were not designed to capture military-specific perspectives. We do not believe this small subset of views substantially shifts the civilian viewpoint, making Table~\ref{tab:compare} a conservative comparison.

\newcommand{\cmark}{\ding{51}} %
\newcommand{\xmark}{\ding{55}} %

\begin{table}[t]
\centering
\small
\caption{Military vs. Civilian Mariner Cyber Perspectives}
\begin{tabular}{lcc}
\toprule
\textbf{Category} & \textbf{Military} & \textbf{Civilian} \\
\midrule
\multicolumn{3}{l}{\textbf{Threats Discussed}} \\
\hspace{4mm} Navigation & \cmark & \cmark \\
\hspace{4mm} Ransomware & \cmark & \cmark \\
\hspace{4mm} Phishing & \cmark & \cmark \\
\hspace{4mm} Communications & \cmark & \cmark \\
\hspace{4mm} FLSS (Fire Life-Safety Systems) & \cmark & \xmark \\
\hspace{4mm} Propulsion & \cmark & \xmark \\
\hspace{4mm} Weapons & \cmark & \xmark \\
\hspace{4mm} Information & \cmark & \xmark \\
\midrule
\multicolumn{3}{l}{\textbf{Perceived Impacts}} \\
\hspace{4mm} Operational Disruption & \cmark & \cmark \\
\hspace{4mm} Economic Loss & \cmark & \cmark \\
\hspace{4mm} Safety Risk & \cmark & \cmark \\
\hspace{4mm} Weapon Hijacking & \cmark & \xmark \\
\hspace{4mm} Geopolitical Escalation & \cmark & \xmark \\
\hspace{4mm} National Security & \cmark & \xmark \\
\hspace{4mm} Public Image Harm & \cmark & \xmark \\
\hspace{4mm} Integrity Loss & \cmark & \xmark \\
\hspace{4mm} Confidentiality Breaches & \cmark & \xmark \\
\midrule
\multicolumn{3}{l}{\textbf{Mitigation Measures}} \\
\hspace{4mm} Basic Cybersecurity Training & \cmark & \cmark \\
\hspace{4mm} System Hardening & \cmark & \xmark \\
\hspace{4mm} OPSEC / EMCON & \cmark & \xmark \\
\hspace{4mm} Specialized Navigation Systems & \cmark & \xmark \\
\hspace{4mm} Cyber Response Teams & \cmark & \xmark \\
\bottomrule
\end{tabular}

\label{tab:compare}
\end{table}

Across both military and civilian maritime domains, improving cybersecurity should be treated as a shared goal of securing the broader maritime environment. While military mariners described a wider range of threats and impact vectors in Table~\ref{tab:compare}, most of the vulnerable systems (e.g., \gls{flss}, Propulsion, Information) and impacts (e.g., public image harm, integrity loss, confidentiality breaches) are common to both sectors. In Section~\ref{subsub:military_civ}, we identified hardening performed across military vessels. We recognize that military operators have access to resources most commercial operators do not, but this is precisely where military experience can help: by pointing to which hardening practices matter the most. The military perspective could help civilian operators prioritize limited resources on the most safety and mission-critical systems. Standardizing cybersecurity assessments in civilian maritime contexts would enable more efficient and targeted investments.

We also recommend the adoption of the \gls{dod}'s Cyber Awareness Challenge, or an international equivalent, for any maritime organization or vessel without annual compliance training. This training, described by nearly all participants and required of all federal employees, is a 60-minute online module that is available, for free, to anyone~\cite{dod_cyber_challenge}. While previous research argues that cybersecurity training must be role-specific to be effective~\cite{raymaker2025sea, flaa2024cybersecurity}, in Section~\ref{subsub:chain_of_command}, we identified a clear understanding of individual cybersecurity responsibility from this training. The training is updated annually to reflect current threats and provides cost-effective, actionable training to establish baseline habits.

\subsection{Future Work}
Several directions remain for future work. First, extending this study to other military branches (e.g., Air Force or Army) would test how far the response model, training norms, and organizational constraints generalize beyond maritime systems. Aviation, in particular, offers a strong comparison point, blending mission-critical systems with adversarial operating environments. Second, studying foreign militaries would show how policy, training pipelines, resource availability, and threat exposure shape operator reasoning, and would help separate broadly shared military practices from those specific to the \gls{us} Third, sharing sanitized versions of the military's cyber hardening and assessment processes could bridge the gap between security work and operator understanding.

\begin{acks}
We thank each mariner who gave their time to support this work. Their firsthand knowledge and insights were invaluable, and this work could not have been possible without these contributions. Of note, any views and conclusions contained herein are those of the author and do not represent the official positions, express or implied, of the \gls{us} Government.
\end{acks}

\bibliographystyle{ACM-Reference-Format}
\bibliography{references}


\begin{thebibliography}{88}


\ifx \showCODEN    \undefined \def \showCODEN     #1{\unskip}     \fi
\ifx \showISBNx    \undefined \def \showISBNx     #1{\unskip}     \fi
\ifx \showISBNxiii \undefined \def \showISBNxiii  #1{\unskip}     \fi
\ifx \showISSN     \undefined \def \showISSN      #1{\unskip}     \fi
\ifx \showLCCN     \undefined \def \showLCCN      #1{\unskip}     \fi
\ifx \shownote     \undefined \def \shownote      #1{#1}          \fi
\ifx \showarticletitle \undefined \def \showarticletitle #1{#1}   \fi
\ifx \showURL      \undefined \def \showURL       {\relax}        \fi
\providecommand\bibfield[2]{#2}
\providecommand\bibinfo[2]{#2}
\providecommand\natexlab[1]{#1}
\providecommand\showeprint[2][]{arXiv:#2}

\bibitem[Abukhousa et~al\mbox{.}(2025)]%
        {abukhousa2025wisdom}
\bibfield{author}{\bibinfo{person}{Emad Abukhousa}, \bibinfo{person}{Syed
  Sohail Feroz~Syed Afroz}, \bibinfo{person}{Fahad Alsaeed},
  \bibinfo{person}{Abdulaziz Qwbaiban}, \bibinfo{person}{Saman Zonouz}, {and}
  \bibinfo{person}{AP~Sakis Meliopoulos}.} \bibinfo{year}{2025}\natexlab{}.
\newblock \showarticletitle{The Wisdom of the Crowd: High-Fidelity
  Classification of Cyber-Attacks and Faults in Power Systems Using Ensemble
  and Machine Learning}. In \bibinfo{booktitle}{\emph{2025 IEEE PES Conference
  on Innovative Smart Grid Technologies-Middle East (ISGT Middle East)}}. IEEE.
\newblock


\bibitem[Adams(2010)]%
        {adams2010shipboard}
\bibfield{author}{\bibinfo{person}{Michael~R Adams}.}
  \bibinfo{year}{2010}\natexlab{}.
\newblock \bibinfo{booktitle}{\emph{Shipboard Bridge Resource Management}}.
\newblock \bibinfo{publisher}{Paradise Cay Publications}.
\newblock


\bibitem[Akgul et~al\mbox{.}(2023)]%
        {akgul2023bug}
\bibfield{author}{\bibinfo{person}{Omer Akgul}, \bibinfo{person}{Taha
  Eghtesad}, \bibinfo{person}{Amit Elazari}, \bibinfo{person}{Omprakash
  Gnawali}, \bibinfo{person}{Jens Grossklags}, \bibinfo{person}{Michelle~L
  Mazurek}, \bibinfo{person}{Daniel Votipka}, {and} \bibinfo{person}{Aron
  Laszka}.} \bibinfo{year}{2023}\natexlab{}.
\newblock \showarticletitle{Bug Hunters' Perspectives on the Challenges and
  Benefits of the Bug Bounty Ecosystem}. In \bibinfo{booktitle}{\emph{32nd
  USENIX Security Symposium (USENIX Security 23)}}.
\newblock


\bibitem[Betz et~al\mbox{.}(2000)]%
        {gps_mcode}
\bibfield{author}{\bibinfo{person}{John~W. Betz}, \bibinfo{person}{Brian
  Barker}, \bibinfo{person}{John Clark}, \bibinfo{person}{Jeffrey~T. Correia},
  \bibinfo{person}{James~T. Gillis}, \bibinfo{person}{Steven Lazar},
  \bibinfo{person}{Kaysi Rehborn}, {and} \bibinfo{person}{John Straton}.}
  \bibinfo{year}{2000}\natexlab{}.
\newblock \bibinfo{booktitle}{\emph{{Overview of the GPS M Code Signal}}}.
\newblock \bibinfo{type}{Technical Report}. \bibinfo{institution}{The MITRE
  Corporation}.
\newblock
\urldef\tempurl%
\url{https://www.mitre.org/sites/default/files/pdf/betz_overview.pdf}
\showURL{%
\tempurl}


\bibitem[Bhatia et~al\mbox{.}(2021)]%
        {bhatia2021evading}
\bibfield{author}{\bibinfo{person}{Rohit Bhatia}, \bibinfo{person}{Vireshwar
  Kumar}, \bibinfo{person}{Khaled Serag}, \bibinfo{person}{Z~Berkay Celik},
  \bibinfo{person}{Mathias Payer}, {and} \bibinfo{person}{Dongyan Xu}.}
  \bibinfo{year}{2021}\natexlab{}.
\newblock \showarticletitle{Evading Voltage-Based Intrusion Detection on
  Automotive CAN}. In \bibinfo{booktitle}{\emph{Network and Distributed Systems
  Security (NDSS) Symposium 2021}}.
\newblock


\bibitem[Bhatti and Humphreys(2017)]%
        {bhatti2017hostile}
\bibfield{author}{\bibinfo{person}{Jahshan Bhatti} {and}
  \bibinfo{person}{Todd~E Humphreys}.} \bibinfo{year}{2017}\natexlab{}.
\newblock \showarticletitle{Hostile Control of Ships via False GPS Signals:
  Demonstration and Detection}.
\newblock \bibinfo{journal}{\emph{NAVIGATION: Journal of the Institute of
  Navigation}}  \bibinfo{volume}{64} (\bibinfo{year}{2017}).
\newblock


\bibitem[{BIMCO and Industry Consortium}(2024)]%
        {bimco2024risk}
\bibfield{author}{\bibinfo{person}{{BIMCO and Industry Consortium}}.}
  \bibinfo{year}{2024}\natexlab{}.
\newblock \bibinfo{title}{{Guidelines on Cyber Security Onboard Ships (Version
  5)}}.
\newblock \bibinfo{howpublished}{Industry guidance document, Maritime Global
  Security}.
\newblock
\urldef\tempurl%
\url{https://www.maritimeglobalsecurity.org/media/g3qlxdaw/2024-11-14-guidelines_on_cyber_security-v5-final.pdf}
\showURL{%
\tempurl}


\bibitem[Birnbach et~al\mbox{.}(2017)]%
        {birnbach2017wi}
\bibfield{author}{\bibinfo{person}{Simon Birnbach}, \bibinfo{person}{Richard
  Baker}, {and} \bibinfo{person}{Ivan Martinovic}.}
  \bibinfo{year}{2017}\natexlab{}.
\newblock \showarticletitle{Wi-fly?: Detecting Privacy Invasion Attacks by
  Consumer Drones}. In \bibinfo{booktitle}{\emph{Network and Distributed
  Systems Security (NDSS) Symposium 2017}}.
\newblock


\bibitem[Bisping et~al\mbox{.}(2024)]%
        {bisping2024wireless}
\bibfield{author}{\bibinfo{person}{Robin Bisping}, \bibinfo{person}{Johannes
  Willbold}, \bibinfo{person}{Martin Strohmeier}, {and}
  \bibinfo{person}{Vincent Lenders}.} \bibinfo{year}{2024}\natexlab{}.
\newblock \showarticletitle{Wireless Signal Injection Attacks on {VSAT}
  Satellite Modems}. In \bibinfo{booktitle}{\emph{33rd USENIX Security
  Symposium (USENIX Security 24)}}.
\newblock


\bibitem[Bottero(2025)]%
        {bottero2025systems}
\bibfield{author}{\bibinfo{person}{Mattia Bottero}.}
  \bibinfo{year}{2025}\natexlab{}.
\newblock \emph{\bibinfo{title}{Systems Engineering Method and Tools for
  Capability-Based Approach in Naval Ships Design}}.
\newblock \bibinfo{thesistype}{Ph.\,D. Dissertation}.
  \bibinfo{school}{Universit{\`a} degli studi di Genova}.
\newblock


\bibitem[Bowditch(1906)]%
        {bowditch1906american}
\bibfield{author}{\bibinfo{person}{Nathaniel Bowditch}.}
  \bibinfo{year}{1906}\natexlab{}.
\newblock \bibinfo{booktitle}{\emph{American Practical Navigator}}.
\newblock Number~9. \bibinfo{publisher}{US Navy Hydrographic Office under the
  authority of the Secretary of the Navy}.
\newblock


\bibitem[Braun and Clarke(2006)]%
        {braun2006using}
\bibfield{author}{\bibinfo{person}{Virginia Braun} {and}
  \bibinfo{person}{Victoria Clarke}.} \bibinfo{year}{2006}\natexlab{}.
\newblock \showarticletitle{Using Thematic Analysis in Psychology}.
\newblock \bibinfo{journal}{\emph{Qualitative research in psychology}}
  (\bibinfo{year}{2006}).
\newblock


\bibitem[Braun and Clarke(2019)]%
        {braun2019reflecting}
\bibfield{author}{\bibinfo{person}{Virginia Braun} {and}
  \bibinfo{person}{Victoria Clarke}.} \bibinfo{year}{2019}\natexlab{}.
\newblock \showarticletitle{Reflecting on reflexive thematic analysis}.
\newblock \bibinfo{journal}{\emph{Qualitative research in sport, exercise and
  health}} (\bibinfo{year}{2019}).
\newblock


\bibitem[Chen et~al\mbox{.}(2024)]%
        {chen2024motivates}
\bibfield{author}{\bibinfo{person}{Xiaowei Chen}, \bibinfo{person}{Sophie
  Doublet}, \bibinfo{person}{Anastasia Sergeeva}, \bibinfo{person}{Gabriele
  Lenzini}, \bibinfo{person}{Vincent Koenig}, {and} \bibinfo{person}{Verena
  Distler}.} \bibinfo{year}{2024}\natexlab{}.
\newblock \showarticletitle{What Motivates and Discourages Employees in
  Phishing Interventions: An Exploration of Expectancy-Value Theory}. In
  \bibinfo{booktitle}{\emph{20th Symposium on Usable Privacy and Security
  (SOUPS 2024)}}.
\newblock


\bibitem[Cutler(2004)]%
        {cutler2004dutton}
\bibfield{author}{\bibinfo{person}{Thomas~J Cutler}.}
  \bibinfo{year}{2004}\natexlab{}.
\newblock \bibinfo{booktitle}{\emph{Dutton's Nautical Navigation}}.
  Vol.~\bibinfo{volume}{15}.
\newblock \bibinfo{publisher}{Naval Institute Press Annapolis}.
\newblock


\bibitem[{CyberPeace Institute}(2022)]%
        {ViasatCaseStudy}
\bibfield{author}{\bibinfo{person}{{CyberPeace Institute}}.}
  \bibinfo{year}{2022}\natexlab{}.
\newblock \bibinfo{title}{Case Study: Viasat}.
\newblock \bibinfo{howpublished}{Cyber Peace Institute: Cyber Conflicts Case
  Study}.
\newblock
\urldef\tempurl%
\url{https://cyberconflicts.cyberpeaceinstitute.org/law-and-policy/cases/viasat}
\showURL{%
\tempurl}


\bibitem[Danielsen et~al\mbox{.}(2021)]%
        {danielsen2021still}
\bibfield{author}{\bibinfo{person}{Brit-Eli Danielsen},
  \bibinfo{person}{Margareta L{\"u}tzh{\"o}ft}, {and} \bibinfo{person}{Thomas
  Porathe}.} \bibinfo{year}{2021}\natexlab{}.
\newblock \showarticletitle{Still Unresolved After All These Years:
  Human-Technology Interaction in the Maritime Domain}. In
  \bibinfo{booktitle}{\emph{International Conference on Applied Human Factors
  and Ergonomics}}. Springer.
\newblock


\bibitem[Das et~al\mbox{.}(2014)]%
        {das2014effect}
\bibfield{author}{\bibinfo{person}{Sauvik Das}, \bibinfo{person}{Tiffany
  Hyun-Jin Kim}, \bibinfo{person}{Laura~A Dabbish}, {and}
  \bibinfo{person}{Jason~I Hong}.} \bibinfo{year}{2014}\natexlab{}.
\newblock \showarticletitle{The Effect of Social Influence on Security
  Sensitivity}. In \bibinfo{booktitle}{\emph{10th Symposium On Usable Privacy
  and Security (SOUPS 2014)}}.
\newblock


\bibitem[Debusmann(2023)]%
        {dod_email_domain}
\bibfield{author}{\bibinfo{person}{Bernd Debusmann, Jr.}}
  \bibinfo{year}{2023}\natexlab{}.
\newblock \bibinfo{title}{{Typo Sends Millions of US Military Emails to Russian
  Ally Mali}}.
\newblock \bibinfo{howpublished}{BBC News}.
\newblock
\urldef\tempurl%
\url{https://www.bbc.com/news/world-us-canada-66226873}
\showURL{%
\tempurl}


\bibitem[Demchak and Tangredi(2025)]%
        {demchak2025cyber}
\bibfield{author}{\bibinfo{person}{Chris~C Demchak} {and}
  \bibinfo{person}{Sam~J Tangredi}.} \bibinfo{year}{2025}\natexlab{}.
\newblock \bibinfo{booktitle}{\emph{Cyber Warfare and Navies: Digital Conflict
  in the Maritime Domain}}.
\newblock \bibinfo{publisher}{Naval Institute Press}.
\newblock


\bibitem[Denning(1987)]%
        {denning1987intrusion}
\bibfield{author}{\bibinfo{person}{Dorothy~E Denning}.}
  \bibinfo{year}{1987}\natexlab{}.
\newblock \showarticletitle{An Intrusion-Detection Model}.
\newblock \bibinfo{journal}{\emph{IEEE Transactions on Software Engineering}}
  (\bibinfo{year}{1987}).
\newblock


\bibitem[{Det Norske Veritas}(2026)]%
        {DNV_RU_SHIP_Pt6Ch2Sec7}
\bibfield{author}{\bibinfo{person}{{Det Norske Veritas}}.}
  \bibinfo{year}{2026}\natexlab{}.
\newblock \bibinfo{title}{{DNV-RU-SHIP} Pt.6 Ch.2 Sec.7: Redundant Propulsion}.
\newblock
\urldef\tempurl%
\url{https://www.dnv.com/rules-standards/}
\showURL{%
\tempurl}


\bibitem[Evripidou et~al\mbox{.}(2023)]%
        {evripidou2023exploring}
\bibfield{author}{\bibinfo{person}{Stefanos Evripidou},
  \bibinfo{person}{Uchenna~D Ani}, \bibinfo{person}{Stephen Hailes}, {and}
  \bibinfo{person}{Jeremy D~McK Watson}.} \bibinfo{year}{2023}\natexlab{}.
\newblock \showarticletitle{Exploring the Security Culture of Operational
  Technology Organisations: the Role of External Consultancy in Overcoming
  Organisational Barriers}. In \bibinfo{booktitle}{\emph{19th Symposium on
  Usable Privacy and Security (SOUPS 2023)}}.
\newblock


\bibitem[Evripidou and Daniel McKendrick~Watson(2024)]%
        {evripidou2024understanding}
\bibfield{author}{\bibinfo{person}{Stefanos Evripidou} {and}
  \bibinfo{person}{Jeremy Daniel McKendrick~Watson}.}
  \bibinfo{year}{2024}\natexlab{}.
\newblock \showarticletitle{Understanding Operational Technology Personnel's
  Mindsets and Their Effect on Cybersecurity Perceptions: A Qualitative Study
  With Operational Technology Cybersecurity Practitioners}. In
  \bibinfo{booktitle}{\emph{Proceedings of the 2024 European Symposium on
  Usable Security}}. ACM.
\newblock


\bibitem[Fenton(2024)]%
        {fenton2024preventing}
\bibfield{author}{\bibinfo{person}{Adam~James Fenton}.}
  \bibinfo{year}{2024}\natexlab{}.
\newblock \showarticletitle{Preventing Catastrophic Cyber-Physical Attacks on
  the Global Maritime Transportation System: a Case Study of Hybrid Maritime
  Security in the Straits of Malacca and Singapore}.
\newblock \bibinfo{journal}{\emph{Journal of Marine Science and Engineering}}
  \bibinfo{volume}{12} (\bibinfo{year}{2024}).
\newblock


\bibitem[Ferazza and Mersinas(2025)]%
        {ferazza2025non}
\bibfield{author}{\bibinfo{person}{Francesco Ferazza} {and}
  \bibinfo{person}{Konstantinos Mersinas}.} \bibinfo{year}{2025}\natexlab{}.
\newblock \showarticletitle{Non-Kinetic Naval Strategy: the Role of Cyber
  Operations in Modern Maritime Conflict}.
\newblock \bibinfo{journal}{\emph{Journal of Cybersecurity}}
  \bibinfo{volume}{11}, \bibinfo{number}{1} (\bibinfo{year}{2025}).
\newblock


\bibitem[Fl{\aa} et~al\mbox{.}(2024)]%
        {flaa2024cybersecurity}
\bibfield{author}{\bibinfo{person}{Lars~Halvdan Fl{\aa}},
  \bibinfo{person}{Christoph~Alexander Thieme}, \bibinfo{person}{Martin~Gilje
  Jaatun}, {and} \bibinfo{person}{Geir~Kjetil Hanssen}.}
  \bibinfo{year}{2024}\natexlab{}.
\newblock \showarticletitle{Cybersecurity Challenges in Industrial Control
  Systems: An Interview Study with Asset Owners in Norway}. In
  \bibinfo{booktitle}{\emph{European Symposium on Research in Computer
  Security}}. Springer.
\newblock


\bibitem[Forson-Adaboh(2022)]%
        {forson2022assessing}
\bibfield{author}{\bibinfo{person}{Kwadwo Forson-Adaboh}.}
  \bibinfo{year}{2022}\natexlab{}.
\newblock \emph{\bibinfo{title}{Assessing Maritime Cyber Security Awareness in
  Navies of the Gulf of Guinea Countries: a Case Study of Ghana}}.
\newblock \bibinfo{thesistype}{Ph.\,D. Dissertation}. \bibinfo{school}{World
  Maritime University}.
\newblock


\bibitem[Fung et~al\mbox{.}(2025)]%
        {fung2025adopting}
\bibfield{author}{\bibinfo{person}{Clement Fung}, \bibinfo{person}{Eric Zeng},
  {and} \bibinfo{person}{Lujo Bauer}.} \bibinfo{year}{2025}\natexlab{}.
\newblock \showarticletitle{Adopting {AI} to Protect Industrial Control
  Systems: Assessing Challenges and Opportunities from the Operators'
  Perspective}. In \bibinfo{booktitle}{\emph{21st Symposium on Usable Privacy
  and Security (SOUPS 2025)}}.
\newblock


\bibitem[Gallardo et~al\mbox{.}(2024)]%
        {gallardo2024interdisciplinary}
\bibfield{author}{\bibinfo{person}{Andrea Gallardo}, \bibinfo{person}{Robert
  Erbes}, \bibinfo{person}{Katya Le~Blanc}, \bibinfo{person}{Lujo Bauer}, {and}
  \bibinfo{person}{Lorrie~Faith Cranor}.} \bibinfo{year}{2024}\natexlab{}.
\newblock \showarticletitle{Interdisciplinary Approaches to Cybervulnerability
  Impact Assessment for Energy Critical Infrastructure}. In
  \bibinfo{booktitle}{\emph{Proceedings of the 2024 CHI Conference on Human
  Factors in Computing Systems}}.
\newblock


\bibitem[Goodall et~al\mbox{.}(2004)]%
        {goodall2004work}
\bibfield{author}{\bibinfo{person}{John Goodall}, \bibinfo{person}{Wayne
  Lutters}, {and} \bibinfo{person}{Anita Komlodi}.}
  \bibinfo{year}{2004}\natexlab{}.
\newblock \showarticletitle{The Work of Intrusion Detection: Rethinking the
  Role of Security Analysts}.
\newblock \bibinfo{journal}{\emph{Americas Conference on Information Systems}}
  (\bibinfo{year}{2004}).
\newblock


\bibitem[Guest et~al\mbox{.}(2020)]%
        {guest2020simple}
\bibfield{author}{\bibinfo{person}{Greg Guest}, \bibinfo{person}{Emily Namey},
  {and} \bibinfo{person}{Mario Chen}.} \bibinfo{year}{2020}\natexlab{}.
\newblock \showarticletitle{A Simple Method to Assess and Report Thematic
  Saturation in Qualitative Research}.
\newblock \bibinfo{journal}{\emph{PloS one}}  \bibinfo{volume}{15}
  (\bibinfo{year}{2020}).
\newblock


\bibitem[Hartley(2012)]%
        {hartley2012economics}
\bibfield{author}{\bibinfo{person}{Keith Hartley}.}
  \bibinfo{year}{2012}\natexlab{}.
\newblock \bibinfo{booktitle}{\emph{The Economics of Defence Policy: A New
  Perspective}}.
\newblock \bibinfo{publisher}{Routledge Studies in Defence and Peace
  Economics}.
\newblock


\bibitem[Hayes and Krippendorff(2007)]%
        {hayes2007answering}
\bibfield{author}{\bibinfo{person}{Andrew~F Hayes} {and} \bibinfo{person}{Klaus
  Krippendorff}.} \bibinfo{year}{2007}\natexlab{}.
\newblock \showarticletitle{Answering the call for a standard reliability
  measure for coding data}.
\newblock \bibinfo{journal}{\emph{Communication methods and measures}}
  \bibinfo{volume}{1} (\bibinfo{year}{2007}).
\newblock


\bibitem[Hu et~al\mbox{.}(2021)]%
        {hu2021automated}
\bibfield{author}{\bibinfo{person}{Shengtuo Hu}, \bibinfo{person}{Qi~Alfred
  Chen}, \bibinfo{person}{Jiachen Sun}, \bibinfo{person}{Yiheng Feng},
  \bibinfo{person}{Z~Morley Mao}, {and} \bibinfo{person}{Henry~X Liu}.}
  \bibinfo{year}{2021}\natexlab{}.
\newblock \showarticletitle{Automated Discovery of Denial-of-Service
  Vulnerabilities in Connected Vehicle Protocols}. In
  \bibinfo{booktitle}{\emph{30th USENIX Security Symposium (USENIX Security
  21)}}.
\newblock


\bibitem[{International Maritime Organization}(2002)]%
        {IMO_SOLAS_V_R19}
\bibfield{author}{\bibinfo{person}{{International Maritime Organization}}.}
  \bibinfo{year}{2002}\natexlab{}.
\newblock \bibinfo{title}{{International Convention for the Safety of Life at
  Sea} Chapter {V}, Regulation 19: Carriage Requirements for Shipborne
  Navigational Systems and Equipment}.
\newblock
\urldef\tempurl%
\url{https://media.liscr.com/marketing/liscr/media/liscr/online\%20library/maritime/solas\%20v_reg19.pdf}
\showURL{%
\tempurl}


\bibitem[{International Maritime Organization}(2017)]%
        {imo2017}
\bibfield{author}{\bibinfo{person}{{International Maritime Organization}}.}
  \bibinfo{year}{2017}\natexlab{}.
\newblock \bibinfo{title}{Resolution MSC.428(98) on Maritime Cyber Risk
  Management in Safety Management Systems}.
\newblock
\urldef\tempurl%
\url{https://wwwcdn.imo.org/localresources/en/OurWork/Security/Documents/Resolution%20MSC.428(98).pdf}
\showURL{%
\tempurl}


\bibitem[{International Maritime Organization}(2025)]%
        {imo2025risk}
\bibfield{author}{\bibinfo{person}{{International Maritime Organization}}.}
  \bibinfo{year}{2025}\natexlab{}.
\newblock \bibinfo{title}{Guidelines on Maritime Cyber Risk Management
  (MSC-FAL.1/Circ.3/Rev.3)}.
\newblock
\urldef\tempurl%
\url{https://wwwcdn.imo.org/localresources/en/OurWork/Facilitation/FAL%20related%20nonmandatory%20documents/MSC-FAL.1-Circ.3-Rev.3.pdf}
\showURL{%
\tempurl}


\bibitem[{International Maritime Organization. Maritime Safety
  Committee}(2022)]%
        {IMO_MSC496_105}
\bibfield{author}{\bibinfo{person}{{International Maritime Organization.
  Maritime Safety Committee}}.} \bibinfo{year}{2022}\natexlab{}.
\newblock \bibinfo{title}{{Resolution MSC.496(105)}: Amendments to the
  International Convention for the Safety of Life at Sea, 1974}.
\newblock
\urldef\tempurl%
\url{https://wwwcdn.imo.org/localresources/en/KnowledgeCentre/IndexofIMOResolutions/MSCResolutions/MSC.496(105).pdf}
\showURL{%
\tempurl}


\bibitem[Jansen et~al\mbox{.}(2021)]%
        {jansen2021trust}
\bibfield{author}{\bibinfo{person}{Kai Jansen}, \bibinfo{person}{Liang Niu},
  \bibinfo{person}{Nian Xue}, \bibinfo{person}{Ivan Martinovic}, {and}
  \bibinfo{person}{Christina P{\"o}pper}.} \bibinfo{year}{2021}\natexlab{}.
\newblock \showarticletitle{Trust the Crowd: Wireless Witnessing to Detect
  Attacks on ADS-B-Based Air-Traffic Surveillance}. In
  \bibinfo{booktitle}{\emph{Network and Distributed Systems Security (NDSS)
  Symposium 2021}}.
\newblock


\bibitem[Jansen et~al\mbox{.}(2017)]%
        {jansen2017localization}
\bibfield{author}{\bibinfo{person}{Kai Jansen}, \bibinfo{person}{Matthias
  Sch{\"a}fer}, \bibinfo{person}{Vincent Lenders}, \bibinfo{person}{Christina
  P{\"o}pper}, {and} \bibinfo{person}{Jens Schmitt}.}
  \bibinfo{year}{2017}\natexlab{}.
\newblock \showarticletitle{Localization of Spoofing Devices Using a
  Large-Scale Air Traffic Surveillance System}. In
  \bibinfo{booktitle}{\emph{Proceedings of the 2017 ACM on Asia conference on
  computer and communications security}}.
\newblock


\bibitem[Jing et~al\mbox{.}(2024)]%
        {jing2024revisiting}
\bibfield{author}{\bibinfo{person}{Pengfei Jing}, \bibinfo{person}{Zhiqiang
  Cai}, \bibinfo{person}{Yingjie Cao}, \bibinfo{person}{Le Yu},
  \bibinfo{person}{Yuefeng Du}, \bibinfo{person}{Wenkai Zhang},
  \bibinfo{person}{Chenxiong Qian}, \bibinfo{person}{Xiapu Luo},
  \bibinfo{person}{Sen Nie}, {and} \bibinfo{person}{Shi Wu}.}
  \bibinfo{year}{2024}\natexlab{}.
\newblock \showarticletitle{Revisiting Automotive Attack Surfaces: A
  Practitioners' Perspective}. In \bibinfo{booktitle}{\emph{2024 IEEE Symposium
  on Security and Privacy (SP)}}. IEEE.
\newblock


\bibitem[{Joint Chiefs of Staff}(2018)]%
        {dod_dictionary}
\bibfield{author}{\bibinfo{person}{{Joint Chiefs of Staff}}.}
  \bibinfo{year}{2018}\natexlab{}.
\newblock \bibinfo{booktitle}{\emph{{DOD Dictionary of Military and Associated
  Terms}}}.
\newblock \bibinfo{type}{{T}echnical {R}eport}. \bibinfo{institution}{U.S.
  Department of Defense}.
\newblock
\urldef\tempurl%
\url{https://apps.dtic.mil/sti/pdfs/ADA542006.pdf}
\showURL{%
\tempurl}


\bibitem[Koch and Golling(2016)]%
        {koch2016weapons}
\bibfield{author}{\bibinfo{person}{Robert Koch} {and} \bibinfo{person}{Mario
  Golling}.} \bibinfo{year}{2016}\natexlab{}.
\newblock \showarticletitle{Weapons Systems and Cyber Security--A Challenging
  Union}. In \bibinfo{booktitle}{\emph{2016 8th International Conference on
  Cyber Conflict (CyCon)}}. IEEE.
\newblock


\bibitem[Li et~al\mbox{.}(2024)]%
        {li2024usability}
\bibfield{author}{\bibinfo{person}{Karen Li}, \bibinfo{person}{Kopo
  Ramokapane}, {and} \bibinfo{person}{Awais Rashid}.}
  \bibinfo{year}{2024}\natexlab{}.
\newblock \showarticletitle{Usability Study of Security Features in
  Programmable Logic Controllers}. In \bibinfo{booktitle}{\emph{Proceedings of
  the 2024 European Symposium on Usable Security}}. ACM.
\newblock


\bibitem[Libicki(2009)]%
        {libicki2009cyberdeterrence}
\bibfield{author}{\bibinfo{person}{Martin~C Libicki}.}
  \bibinfo{year}{2009}\natexlab{}.
\newblock \bibinfo{booktitle}{\emph{Cyberdeterrence and Cyberwar}}.
\newblock \bibinfo{publisher}{RAND Corporation}.
\newblock


\bibitem[Liu et~al\mbox{.}(2021)]%
        {liu2021stars}
\bibfield{author}{\bibinfo{person}{Shinan Liu}, \bibinfo{person}{Xiang Cheng},
  \bibinfo{person}{Hanchao Yang}, \bibinfo{person}{Yuanchao Shu},
  \bibinfo{person}{Xiaoran Weng}, \bibinfo{person}{Ping Guo},
  \bibinfo{person}{Kexiong~Curtis Zeng}, \bibinfo{person}{Gang Wang}, {and}
  \bibinfo{person}{Yaling Yang}.} \bibinfo{year}{2021}\natexlab{}.
\newblock \showarticletitle{Stars Can Tell: A Robust Method to Defend Against
  {GPS} Spoofing Attacks Using Off-the-Shelf Chipset}. In
  \bibinfo{booktitle}{\emph{30th USENIX Security Symposium (USENIX Security
  21)}}.
\newblock


\bibitem[Lonergan and Schneider(2023)]%
        {lonergan2023power}
\bibfield{author}{\bibinfo{person}{Erica~D Lonergan} {and}
  \bibinfo{person}{Jacquelyn Schneider}.} \bibinfo{year}{2023}\natexlab{}.
\newblock \showarticletitle{The Power of Beliefs in US Cyber Strategy: The
  Evolving Role of Deterrence, Norms, and Escalation}.
\newblock \bibinfo{journal}{\emph{Journal of Cybersecurity}}
  \bibinfo{volume}{9}, \bibinfo{number}{1} (\bibinfo{year}{2023}).
\newblock


\bibitem[Longo et~al\mbox{.}(2023)]%
        {longo2023attacking}
\bibfield{author}{\bibinfo{person}{Giacomo Longo}, \bibinfo{person}{Enrico
  Russo}, \bibinfo{person}{Alessandro Armando}, {and} \bibinfo{person}{Alessio
  Merlo}.} \bibinfo{year}{2023}\natexlab{}.
\newblock \showarticletitle{Attacking (and Defending) the Maritime Radar
  System}.
\newblock \bibinfo{journal}{\emph{IEEE Transactions on Information Forensics
  and Security}}  \bibinfo{volume}{18} (\bibinfo{year}{2023}).
\newblock


\bibitem[Lundberg et~al\mbox{.}(2014)]%
        {lundberg2014security}
\bibfield{author}{\bibinfo{person}{Devin Lundberg}, \bibinfo{person}{Brown
  Farinholt}, \bibinfo{person}{Edward Sullivan}, \bibinfo{person}{Ryan Mast},
  \bibinfo{person}{Stephen Checkoway}, \bibinfo{person}{Stefan Savage},
  \bibinfo{person}{Alex~C Snoeren}, {and} \bibinfo{person}{Kirill Levchenko}.}
  \bibinfo{year}{2014}\natexlab{}.
\newblock \showarticletitle{On the Security of Mobile Cockpit Information
  Systems}. In \bibinfo{booktitle}{\emph{Proceedings of the 2014 ACM SIGSAC
  Conference on Computer and Communications Security}}.
\newblock


\bibitem[Maxam~III and Davis(2024)]%
        {maxam2024interview}
\bibfield{author}{\bibinfo{person}{William~P Maxam~III} {and}
  \bibinfo{person}{James~C Davis}.} \bibinfo{year}{2024}\natexlab{}.
\newblock \showarticletitle{An Interview Study on Third-Party Cyber Threat
  Hunting Processes in the {US} Department of Homeland Security}. In
  \bibinfo{booktitle}{\emph{33rd USENIX Security Symposium (USENIX Security
  24)}}.
\newblock


\bibitem[{Naval History and Heritage Command}(2006)]%
        {navy_propulsion}
\bibfield{author}{\bibinfo{person}{{Naval History and Heritage Command}}.}
  \bibinfo{year}{2006}\natexlab{}.
\newblock \bibinfo{title}{{Navy Ship Propulsion Technologies: Options for
  Reducing Oil Use}}.
\newblock
\urldef\tempurl%
\url{https://www.history.navy.mil/content/dam/nhhc/research/library/online-reading-room/technology/images/navyshippropulsion/navyshippropulsiontech.pdf}
\showURL{%
\tempurl}


\bibitem[Nganga et~al\mbox{.}(2024)]%
        {nganga2024enabling}
\bibfield{author}{\bibinfo{person}{Allan Nganga}, \bibinfo{person}{Joel
  Scanlan}, \bibinfo{person}{Margareta L{\"u}tzh{\"o}ft}, {and}
  \bibinfo{person}{Steven Mallam}.} \bibinfo{year}{2024}\natexlab{}.
\newblock \showarticletitle{Enabling Cyber Resilient Shipping Through Maritime
  Security Operation Center Adoption: A Human Factors Perspective}.
\newblock \bibinfo{journal}{\emph{Applied Ergonomics}}  \bibinfo{volume}{119}
  (\bibinfo{year}{2024}).
\newblock


\bibitem[{Northrop Grumman}(2023)]%
        {navy_imu}
\bibfield{author}{\bibinfo{person}{{Northrop Grumman}}.}
  \bibinfo{year}{2023}\natexlab{}.
\newblock \bibinfo{title}{{Northrop Grumman to Produce New Maritime Navigation
  Sensor for U.S. Navy}}.
\newblock
\urldef\tempurl%
\url{https://news.northropgrumman.com/navigation-systems/northrop-grumman-to-produce-new-maritime-navigation-sensor-for-u-s-navy}
\showURL{%
\tempurl}


\bibitem[Pavur et~al\mbox{.}(2020)]%
        {pavur2020tale}
\bibfield{author}{\bibinfo{person}{James Pavur}, \bibinfo{person}{Daniel
  Moser}, \bibinfo{person}{Martin Strohmeier}, \bibinfo{person}{Vincent
  Lenders}, {and} \bibinfo{person}{Ivan Martinovic}.}
  \bibinfo{year}{2020}\natexlab{}.
\newblock \showarticletitle{A Tale of Sea and Sky on the Security of Maritime
  VSAT Communications}. In \bibinfo{booktitle}{\emph{2020 IEEE Symposium on
  Security and Privacy (SP)}}. IEEE.
\newblock


\bibitem[Pickren et~al\mbox{.}(2024)]%
        {pickren2024release}
\bibfield{author}{\bibinfo{person}{Ryan Pickren}, \bibinfo{person}{Animesh
  Chhotaray}, \bibinfo{person}{Frank Li}, \bibinfo{person}{Saman Zonouz}, {and}
  \bibinfo{person}{Raheem Beyah}.} \bibinfo{year}{2024}\natexlab{}.
\newblock \showarticletitle{Release the Hounds! Automated Inference and
  Empirical Security Evaluation of Field-Deployed PLCs Using Active Network
  Data}. In \bibinfo{booktitle}{\emph{Proceedings of the 2024 on ACM SIGSAC
  Conference on Computer and Communications Security}}.
\newblock


\bibitem[Progoulakis et~al\mbox{.}(2021)]%
        {progoulakis2021cyber}
\bibfield{author}{\bibinfo{person}{Iosif Progoulakis}, \bibinfo{person}{Paul
  Rohmeyer}, {and} \bibinfo{person}{Nikitas Nikitakos}.}
  \bibinfo{year}{2021}\natexlab{}.
\newblock \showarticletitle{Cyber Physical Systems Security for Maritime
  Assets}.
\newblock \bibinfo{journal}{\emph{Journal of Marine Science and Engineering}}
  \bibinfo{volume}{9} (\bibinfo{year}{2021}).
\newblock


\bibitem[Rajaram et~al\mbox{.}(2022)]%
        {rajaram2022guidelines}
\bibfield{author}{\bibinfo{person}{Priyanga Rajaram}, \bibinfo{person}{Mark
  Goh}, {and} \bibinfo{person}{Jianying Zhou}.}
  \bibinfo{year}{2022}\natexlab{}.
\newblock \showarticletitle{Guidelines for Cyber Risk Management in Shipboard
  Operational Technology Systems}. In \bibinfo{booktitle}{\emph{Journal of
  Physics: Conference Series}}, Vol.~\bibinfo{volume}{2311}. IOP Publishing.
\newblock


\bibitem[Ranganathan et~al\mbox{.}(2016)]%
        {ranganathan2016spree}
\bibfield{author}{\bibinfo{person}{Aanjhan Ranganathan},
  \bibinfo{person}{Hildur {\'O}lafsd{\'o}ttir}, {and} \bibinfo{person}{Srdjan
  Capkun}.} \bibinfo{year}{2016}\natexlab{}.
\newblock \showarticletitle{SPREE: A Spoofing Resistant GPS Receiver}. In
  \bibinfo{booktitle}{\emph{Proceedings of the 22nd Annual International
  Conference on Mobile Computing and Networking}}.
\newblock


\bibitem[Raymaker et~al\mbox{.}(2025)]%
        {raymaker2025sea}
\bibfield{author}{\bibinfo{person}{Anna Raymaker}, \bibinfo{person}{Akshaya
  Kumar}, \bibinfo{person}{Miuyin~Yong Wong}, \bibinfo{person}{Ryan Pickren},
  \bibinfo{person}{Animesh Chhotaray}, \bibinfo{person}{Frank Li},
  \bibinfo{person}{Saman Zonouz}, {and} \bibinfo{person}{Raheem Beyah}.}
  \bibinfo{year}{2025}\natexlab{}.
\newblock \showarticletitle{A Sea of Cyber Threats: Maritime Cybersecurity from
  the Perspective of Mariners}. In \bibinfo{booktitle}{\emph{Proceedings of the
  2025 ACM SIGSAC Conference on Computer and Communications Security}}.
\newblock


\bibitem[Reddy(2025)]%
        {reddy2025cyber}
\bibfield{author}{\bibinfo{person}{Rajender~Pell Reddy}.}
  \bibinfo{year}{2025}\natexlab{}.
\newblock \showarticletitle{Cyber Warfare: National Security Implications and
  Strategic Defense Mechanisms}.
\newblock \bibinfo{journal}{\emph{International Journal of Computer Trends and
  Technology (IJCTT)}}  \bibinfo{volume}{73} (\bibinfo{year}{2025}).
\newblock


\bibitem[Sahin et~al\mbox{.}(2023)]%
        {sahin2023investigating}
\bibfield{author}{\bibinfo{person}{Sena Sahin}, \bibinfo{person}{Suood
  Al~Roomi}, \bibinfo{person}{Tara Poteat}, {and} \bibinfo{person}{Frank Li}.}
  \bibinfo{year}{2023}\natexlab{}.
\newblock \showarticletitle{Investigating the Password Policy Practices of
  Website Administrators}. In \bibinfo{booktitle}{\emph{2023 IEEE Symposium on
  Security and Privacy (SP)}}. IEEE.
\newblock


\bibitem[Sasaki et~al\mbox{.}(2022)]%
        {sasaki2022exposed}
\bibfield{author}{\bibinfo{person}{Takayuki Sasaki}, \bibinfo{person}{Akira
  Fujita}, \bibinfo{person}{Carlos~H Ga{\~n}{\'a}n}, \bibinfo{person}{Michel
  van Eeten}, \bibinfo{person}{Katsunari Yoshioka}, {and}
  \bibinfo{person}{Tsutomu Matsumoto}.} \bibinfo{year}{2022}\natexlab{}.
\newblock \showarticletitle{Exposed Infrastructures: Discovery, Attacks and
  Remediation of Insecure ICS Remote Management Devices}. In
  \bibinfo{booktitle}{\emph{2022 IEEE Symposium on Security and Privacy (SP)}}.
  IEEE.
\newblock


\bibitem[Sathaye et~al\mbox{.}(2022)]%
        {sathaye2022semperfi}
\bibfield{author}{\bibinfo{person}{Harshad Sathaye}, \bibinfo{person}{Gerald
  LaMountain}, \bibinfo{person}{Pau Closas}, {and} \bibinfo{person}{Aanjhan
  Ranganathan}.} \bibinfo{year}{2022}\natexlab{}.
\newblock \showarticletitle{Semperfi: Anti-Spoofing GPS Receiver for UAVs}. In
  \bibinfo{booktitle}{\emph{Network and Distributed Systems Security (NDSS)
  Symposium 2022}}.
\newblock


\bibitem[Singer et~al\mbox{.}(2023)]%
        {singer2023shedding}
\bibfield{author}{\bibinfo{person}{Brian Singer}, \bibinfo{person}{Amritanshu
  Pandey}, \bibinfo{person}{Shimiao Li}, \bibinfo{person}{Lujo Bauer},
  \bibinfo{person}{Craig Miller}, \bibinfo{person}{Lawrence Pileggi}, {and}
  \bibinfo{person}{Vyas Sekar}.} \bibinfo{year}{2023}\natexlab{}.
\newblock \showarticletitle{Shedding Light on Inconsistencies in Grid
  Cybersecurity: Disconnects and Recommendations}. In
  \bibinfo{booktitle}{\emph{2023 IEEE Symposium on Security and Privacy (SP)}}.
  IEEE.
\newblock


\bibitem[Srivastava and Hopwood(2009)]%
        {srivastava2009practical}
\bibfield{author}{\bibinfo{person}{Prachi Srivastava} {and}
  \bibinfo{person}{Nick Hopwood}.} \bibinfo{year}{2009}\natexlab{}.
\newblock \showarticletitle{A practical iterative framework for qualitative
  data analysis}.
\newblock \bibinfo{journal}{\emph{International Journal of Qualitative
  Methods}} \bibinfo{volume}{8}, \bibinfo{number}{1} (\bibinfo{year}{2009}),
  \bibinfo{pages}{76--84}.
\newblock


\bibitem[Stenden(2016)]%
        {usn2016spoofing}
\bibfield{author}{\bibinfo{person}{NHL Stenden}.}
  \bibinfo{year}{2016}\natexlab{}.
\newblock \bibinfo{title}{{US Navy Ships Hit by GPS Spoofing in Persian Gulf,
  Iran}}.
\newblock \bibinfo{howpublished}{Maritime Cyber Attack Database (MCAD)}.
\newblock
\urldef\tempurl%
\url{https://www.maritimecybersecurity.nl/incident/yv0601Xmgn}
\showURL{%
\tempurl}


\bibitem[Tibaldo et~al\mbox{.}(2025)]%
        {tibaldo2025gnss}
\bibfield{author}{\bibinfo{person}{Christopher Tibaldo},
  \bibinfo{person}{Harshad Sathaye}, \bibinfo{person}{Giovanni Camurati}, {and}
  \bibinfo{person}{Srdjan Capkun}.} \bibinfo{year}{2025}\natexlab{}.
\newblock \showarticletitle{{GNSS-WASP: GNSS Wide Area SPoofing}}. In
  \bibinfo{booktitle}{\emph{34th USENIX Security Symposium (USENIX Security
  25)}}.
\newblock


\bibitem[Tran et~al\mbox{.}(2021)]%
        {tran2021marine}
\bibfield{author}{\bibinfo{person}{Ky Tran}, \bibinfo{person}{Sid Keene},
  \bibinfo{person}{Erik Fretheim}, {and} \bibinfo{person}{Michail
  Tsikerdekis}.} \bibinfo{year}{2021}\natexlab{}.
\newblock \showarticletitle{Marine Network Protocols and Security Risks}.
\newblock \bibinfo{journal}{\emph{Journal of Cybersecurity and Privacy}}
  \bibinfo{volume}{1}, \bibinfo{number}{2} (\bibinfo{year}{2021}).
\newblock


\bibitem[{United States Coast Guard}(2026)]%
        {uscg_tours}
\bibfield{author}{\bibinfo{person}{{United States Coast Guard}}.}
  \bibinfo{year}{2026}\natexlab{}.
\newblock \bibinfo{title}{Afloat Officer Careers}.
\newblock
\urldef\tempurl%
\url{https://www.gocoastguard.com/careers/officer/afloat}
\showURL{%
\tempurl}


\bibitem[{United States Navy}(2026)]%
        {navy_weapons}
\bibfield{author}{\bibinfo{person}{{United States Navy}}.}
  \bibinfo{year}{2026}\natexlab{}.
\newblock \bibinfo{title}{{U.S. Navy Weapon Systems Fact Files}}.
\newblock
\urldef\tempurl%
\url{https://www.navy.mil/Resources/Fact-Files/}
\showURL{%
\tempurl}


\bibitem[{U.S. Coast Guard}(2025)]%
        {uscg_paths}
\bibfield{author}{\bibinfo{person}{{U.S. Coast Guard}}.}
  \bibinfo{year}{2025}\natexlab{}.
\newblock \bibinfo{title}{{Officer Opportunities: Becoming a Coast Guard
  Officer}}.
\newblock
\urldef\tempurl%
\url{https://www.gocoastguard.com/careers/officer}
\showURL{%
\tempurl}


\bibitem[{U.S. Coast Guard}(2026)]%
        {uscg_ood}
\bibfield{author}{\bibinfo{person}{{U.S. Coast Guard}}.}
  \bibinfo{year}{2026}\natexlab{}.
\newblock \bibinfo{title}{{Underway OOD Military Occupational Classification}}.
\newblock \bibinfo{howpublished}{{U.S. Coast Guard}}.
\newblock
\urldef\tempurl%
\url{https://www.cool.osd.mil/uscg/moc/index.html?moc=cg_underway_ood\&tab=overview}
\showURL{%
\tempurl}


\bibitem[{U.S. Department of Defense}({[n.\,d.]})]%
        {dod_cyber_challenge}
\bibfield{author}{\bibinfo{person}{{U.S. Department of Defense}}.}
  \bibinfo{year}{[n.\,d.]}\natexlab{}.
\newblock \bibinfo{title}{Cyber Awareness Challenge}.
\newblock
\urldef\tempurl%
\url{https://www.cyber.mil/cyber-awareness-challenge}
\showURL{%
\tempurl}


\bibitem[{U.S. Department of Defense}(2023)]%
        {DoD2023Demographics}
\bibfield{author}{\bibinfo{person}{{U.S. Department of Defense}}.}
  \bibinfo{year}{2023}\natexlab{}.
\newblock \bibinfo{booktitle}{\emph{2023 Demographics: Profile of the Military
  Community}}.
\newblock \bibinfo{type}{{T}echnical {R}eport}. \bibinfo{institution}{Office of
  the Deputy Assistant Secretary of Defense for Military Community and Family
  Policy}.
\newblock
\urldef\tempurl%
\url{https://download.militaryonesource.mil/12038/MOS/Reports/2023-demographics-report.pdf}
\showURL{%
\tempurl}


\bibitem[{U.S. Department of Homeland Security}(2026)]%
        {cg_mission}
\bibfield{author}{\bibinfo{person}{{U.S. Department of Homeland Security}}.}
  \bibinfo{year}{2026}\natexlab{}.
\newblock \bibinfo{title}{Missions}.
\newblock
\urldef\tempurl%
\url{https://www.history.uscg.mil/Home/Missions/}
\showURL{%
\tempurl}


\bibitem[{U.S. Department of the Navy}(2025)]%
        {usn_paths}
\bibfield{author}{\bibinfo{person}{{U.S. Department of the Navy}}.}
  \bibinfo{year}{2025}\natexlab{}.
\newblock \bibinfo{title}{{Paths to Becoming a Naval Officer}}.
\newblock
\urldef\tempurl%
\url{https://www.navy.com/joining/ways-to-join/officer}
\showURL{%
\tempurl}


\bibitem[{U.S. Department of the Navy}(2026)]%
        {navy_mission}
\bibfield{author}{\bibinfo{person}{{U.S. Department of the Navy}}.}
  \bibinfo{year}{2026}\natexlab{}.
\newblock \bibinfo{title}{Mission}.
\newblock
\urldef\tempurl%
\url{https://www.navy.mil/about/mission/}
\showURL{%
\tempurl}


\bibitem[{U.S. Navy}(2023)]%
        {navy_wifi}
\bibfield{author}{\bibinfo{person}{{U.S. Navy}}.}
  \bibinfo{year}{2023}\natexlab{}.
\newblock \bibinfo{title}{{Command Investigation into the Unauthorized Wi-Fi
  Device Found Onboard USS Manchester (LCS 14) Gold}}.
\newblock
\urldef\tempurl%
\url{https://www.surfpac.navy.mil/Portals/54/Documents/CNSP/FOIA/Reading-Room/MCH(G)%20Wi-FI%20CI%20w%20Endorsements_Redacted.pdf?ver=QXQ-c3MA2jFvFlv_o4SA6w%3D%3D}
\showURL{%
\tempurl}


\bibitem[{U.S. Navy Personnel Command}(2010)]%
        {usn_tours}
\bibfield{author}{\bibinfo{person}{{U.S. Navy Personnel Command}}.}
  \bibinfo{year}{2010}\natexlab{}.
\newblock \bibinfo{title}{{MILPERSMAN} 1306-101: Enlisted Assignment System
  (Change 33)}.
\newblock
\urldef\tempurl%
\url{https://www.mynavyhr.navy.mil/Portals/55/Reference/MILPERSMAN/1000/1300Assignment/1306-101.pdf}
\showURL{%
\tempurl}


\bibitem[{Viasat, Inc.}(2022)]%
        {ViasatOverview}
\bibfield{author}{\bibinfo{person}{{Viasat, Inc.}}}
  \bibinfo{year}{2022}\natexlab{}.
\newblock \bibinfo{title}{{KA-SAT Network cyber attack overview}}.
\newblock
\urldef\tempurl%
\url{https://www.viasat.com/perspectives/corporate/2022/ka-sat-network-cyber-attack-overview/}
\showURL{%
\tempurl}


\bibitem[Von~Brock et~al\mbox{.}(2026)]%
        {osf_repo}
\bibfield{author}{\bibinfo{person}{Ryan Von~Brock}, \bibinfo{person}{Anna
  Raymaker}, \bibinfo{person}{Animesh Chhotaray}, \bibinfo{person}{Frank Li},
  \bibinfo{person}{Saman Zonouz}, {and} \bibinfo{person}{Raheem Beyah}.}
  \bibinfo{year}{2026}\natexlab{}.
\newblock \bibinfo{title}{Cybersecurity with Military Mariners}.
\newblock
\urldef\tempurl%
\url{https://osf.io/7t4wk/overview?view_only=2b1404b66f4d46f8803966b66aa3af31}
\showURL{%
\tempurl}


\bibitem[Votipka et~al\mbox{.}(2018)]%
        {votipka2018hackers}
\bibfield{author}{\bibinfo{person}{Daniel Votipka}, \bibinfo{person}{Rock
  Stevens}, \bibinfo{person}{Elissa Redmiles}, \bibinfo{person}{Jeremy Hu},
  {and} \bibinfo{person}{Michelle Mazurek}.} \bibinfo{year}{2018}\natexlab{}.
\newblock \showarticletitle{Hackers vs. Testers: A Comparison of Software
  Vulnerability Discovery Processes}. In \bibinfo{booktitle}{\emph{2018 IEEE
  Symposium on Security and Privacy (SP)}}. IEEE.
\newblock


\bibitem[Wash et~al\mbox{.}(2021)]%
        {wash2021knowledge}
\bibfield{author}{\bibinfo{person}{Rick Wash}, \bibinfo{person}{Norbert
  Nthala}, {and} \bibinfo{person}{Emilee Rader}.}
  \bibinfo{year}{2021}\natexlab{}.
\newblock \showarticletitle{Knowledge and Capabilities that Non-Expert Users
  Bring to Phishing Detection}. In \bibinfo{booktitle}{\emph{17th Symposium on
  Usable Privacy and Security (SOUPS 2021)}}.
\newblock


\bibitem[Wen et~al\mbox{.}(2020)]%
        {wen2020plug}
\bibfield{author}{\bibinfo{person}{Haohuang Wen}, \bibinfo{person}{Qi~Alfred
  Chen}, {and} \bibinfo{person}{Zhiqiang Lin}.}
  \bibinfo{year}{2020}\natexlab{}.
\newblock \showarticletitle{{Plug-N-Pwned: Comprehensive Vulnerability Analysis
  of OBD-II Dongles as a New Over-the-Air Attack Surface in Automotive IoT}}.
  In \bibinfo{booktitle}{\emph{29th USENIX Security Symposium (USENIX Security
  20)}}.
\newblock


\bibitem[Whittaker(2023)]%
        {dod_email_leak}
\bibfield{author}{\bibinfo{person}{Zack Whittaker}.}
  \bibinfo{year}{2023}\natexlab{}.
\newblock \bibinfo{title}{{Sensitive US Military Emails Spill Online}}.
\newblock \bibinfo{howpublished}{TechCrunch}.
\newblock
\urldef\tempurl%
\url{https://techcrunch.com/2023/02/21/sensitive-united-states-military-emails-spill-online/}
\showURL{%
\tempurl}


\bibitem[Wong et~al\mbox{.}(2024)]%
        {wong2024comparing}
\bibfield{author}{\bibinfo{person}{Miuyin~Yong Wong}, \bibinfo{person}{Matthew
  Landen}, \bibinfo{person}{Frank Li}, \bibinfo{person}{Fabian Monrose}, {and}
  \bibinfo{person}{Mustaque Ahamad}.} \bibinfo{year}{2024}\natexlab{}.
\newblock \showarticletitle{Comparing Malware Evasion Theory with Practice:
  Results from Interviews with Expert Analysts}. In
  \bibinfo{booktitle}{\emph{20th Symposium on Usable Privacy and Security
  (SOUPS 2024)}}.
\newblock


\bibitem[Xue et~al\mbox{.}(2022)]%
        {xue2022said}
\bibfield{author}{\bibinfo{person}{Lei Xue}, \bibinfo{person}{Yangyang Liu},
  \bibinfo{person}{Tianqi Li}, \bibinfo{person}{Kaifa Zhao},
  \bibinfo{person}{Jianfeng Li}, \bibinfo{person}{Le Yu},
  \bibinfo{person}{Xiapu Luo}, \bibinfo{person}{Yajin Zhou}, {and}
  \bibinfo{person}{Guofei Gu}.} \bibinfo{year}{2022}\natexlab{}.
\newblock \showarticletitle{{SAID}: State-Aware Defense Against Injection
  Attacks on in-Vehicle Network}. In \bibinfo{booktitle}{\emph{31st USENIX
  Security Symposium (USENIX Security 22)}}.
\newblock


\end{thebibliography}

\appendix
\section{Open Science}
To support open science, we made our final codebook available at an \gls{OSF} repository~\cite{osf_repo}. The final codebook includes all codes (organized by question), a definition for each, and inclusion and exclusion criteria used to code participant responses for each question. The separate themes file contains all overarching themes and sub-themes, their definitions, and the positive and negative codes that contributed to each theme.
\section{Ethical Considerations}

\paragraph{Stakeholders:}
This study involves multiple stakeholders. The primary ones are active-duty and retired military mariners who participated in the study, whose professional experiences in the military expose them to sensitive operational details and classified contexts. Secondary stakeholders include the USN and USCG as organizations responsible for maritime defense and cybersecurity. Civilian maritime operators and policymakers may also draw lessons from this work. The broader research and cybersecurity communities are also stakeholders, as this study contributes empirical evidence to topics of cybersecurity preparation in military-adjacent domains.

\paragraph{Impacts:}
The primary potential risk to participants is the inadvertent disclosure of sensitive or operationally revealing information that could affect individual careers, organizational security, or even national defense. There is also a risk that findings could be misinterpreted as identifying specific weaknesses in military cyber defenses or could be overgeneralized beyond their intended scope. Potential benefits include a better understanding of how military mariners conceptualize and respond to cyber threats, which may inform safer training, practices, policy discussions, and future research in military and civilian maritime contexts.

\paragraph{Mitigations:}
This study was approved by our institution’s IRB. All participants agreed to an approved consent form before their data was collected in a survey or interview. Participation was voluntary, and participants could decline to answer any question or withdraw answers at any time. Lastly, the study team member conducting interviews explicitly reminded participants that their role was as an academic researcher and that interviews were not being conducted in a military context.

Because the inadvertent disclosure of sensitive information could affect careers, special care was taken to anonymize participant responses and remove potentially sensitive operational details. Identifying information, asset types, locations, and timelines were excluded or generalized. After transcription and screening, all audio recordings were destroyed. To further protect our participants, interview Question \#1 was excluded from our OSF repository ~\cite{osf_repo}. 

Further, because the USN and USCG defend against nation-state cyber threats, no specific defenses, technical configurations, or operational procedures were discussed or reported. The study avoided eliciting or publishing details about platform-specific capabilities or vulnerabilities. Participants were given an example ship configuration to reference throughout the scenario-based questions. Described challenges or gaps were framed at human, organizational, or training levels rather than as actionable technical weaknesses. The research team assessed that the human-oriented findings presented are already mitigated in practice by layered organizational controls and do not materially increase adversarial capability.

\paragraph{Decision:}
The research team determined that the societal and scientific value of understanding military mariners’ cybersecurity perceptions and responses outweighs the mitigated risks. By focusing on operator mental models, organizational norms, and training influences, rather than technical vulnerabilities, this work aims to contribute to safer system design and training without compromising operational security. We concluded that the study meets ethical standards for research involving human subjects in sensitive security contexts.

\section{Military Background} \label{app:military_defs}

Military personnel are divided by rank and rate. Officers are commissioned leaders responsible for command and operational decision-making. Enlisted personnel are lower in rank but hold more technical, hands-on roles. They are classified by rates, which indicate operational specialties (e.g. \gls{ET}, \gls{IT})~\cite{dod_dictionary}. Ranks are written as the letter "O" or "E," followed by a number, denoting higher seniority (e.g. O5 $>$ O2).

\section{Survey Questions} \label{app:survey_qs}

\begin{enumerate}[noitemsep]
    \item Indicate your branch of service:
    \item What is your rank?
    \item How old are you?
    \begin{itemize}
        \item ages
    \end{itemize}
    \item What is your gender?
    \item Are you currently serving aboard a ship?
    \begin{itemize}
        \item If not, how recently have you served in an afloat assignment?
    \end{itemize}
    \item In total, how many years have you served aboard military vessels?
    \item How large is the ship you last served on, in tons?
    \item What is the crew size of the last ship you served on?
    \item Which of the following roles have you held? Select all that apply.
     \begin{itemize}
        \item Deck Watch Officer
        \item Engineering Officer in Training
        \item Assistant Command Security Officer
        \item Command Security Officer
        \item Assistant Engineering Officer
        \item Engineering Officer
        \item Assistant Operations Officer
        \item Operations Officer
        \item Executive Officer
        \item Commanding Officer
        \item Other, please specify:
    \end{itemize}
    \item What operations have your vessels conducted? Select all that apply.
    \begin{itemize}
        \item National security/defense
        \item Search and rescue
        \item Counter drug operations
        \item Alien migrant interdiction operations
        \item Maritime law enforcement
        \item Aids to navigation
        \item Supply
        \item Training
        \item Other, please specify:
    \end{itemize}
    \item What flag state(s) have you sailed under?
\end{enumerate}

\section{Interview Questions} \label{app:interview_qs}

\begin{enumerate}[noitemsep]
    \item What was your pathway to becoming an [insert title]?
    \item What experiences or factors have most shaped your personal view of cybersecurity?
    \item From your understanding, how can cyber-attacks affect ships?
    \item What indicators would you use to identify a cyber-attack?

    \item What military-specific cybersecurity training have you received? 
    \begin{itemize}
        \item How relevant did this training feel to your shipboard duties?
    \end{itemize}
    \item Did your cybersecurity training differentiate between IT systems like administrative networks and others such as propulsion and navigation equipment?
    \item Have you received cybersecurity training tailored to your unit’s mission?
    \item Compared to physical threat response training (such as ATFP or damage control), how would you describe your cybersecurity training?
    \begin{itemize}
        \item How did the frequency of these trainings compare? 
        \item What about the seriousness? 
        \item And, lastly, how would you compare their effectiveness?
    \end{itemize}
    \item How comfortable are you navigating with a complete loss of GPS?
    \begin{itemize}
        \item Does this change in coastal waters compared to the open ocean?
    \end{itemize}
    \item How did your shipboard billet(s) shape your cybersecurity responsibilities?
    \item What aspects of military life make it harder, or easier, to care about cybersecurity?
    \item In your opinion, what is the greatest indicator of cybersecurity expertise?
    \item How does rank influence credibility when reporting or escalating cybersecurity concerns?
    \item How do you distinguish between IT support and cybersecurity?
    \begin{itemize}
        \item How have these roles been separated at your units?
    \end{itemize}
    \item I see from the survey that you’ve conducted several different missions afloat. How has each vessel’s cybersecurity culture changed across different missions?
    \item Do you see differences in cyber readiness between afloat and ashore units?
    \item Do you think military vessels are more resilient to cyber threats compared to merchant vessels? Why or why not?
    \item What do you believe are the most significant cybersecurity vulnerabilities, if any, aboard modern vessels, either military or merchant ships?
    \item What do you think are the most significant consequences of a cyber-attack…
    \begin{itemize}
        \item against a merchant vessel?
        \item And what about against a military vessel? 
    \end{itemize}
    \item What cyber behaviors or practices have been encouraged or discouraged by your command(s)?
    \item In your view, where does responsibility for cybersecurity begin, and how does it flow through the organization?
    \item For the first scenario, you are underway in clear weather, near the coast. Your GPS suddenly indicates an incorrect position that contradicts radar returns and visual navigation cues.
    \begin{itemize}
        \item What immediate actions would you take?
        \item What indicators would you assess to determine the source of the anomaly?
    \end{itemize}
    \item For the second scenario, you are underway in a high-traffic area. The ship’s engine control system begins responding erratically, reducing propulsion without any mechanical warning.
    \begin{itemize}
        \item What immediate actions would you take?
        \item What indicators would you assess to determine the source of the anomaly?
    \end{itemize}
    \item For the last scenario, you are transiting a congested channel at night. All of the ship’s electronic navigation systems suddenly go offline. The screens are black. The power plant and rudder remain responsive.
    \begin{itemize}
        \item What immediate actions would you take?
        \item What indicators would you assess to determine the source of the anomaly?
    \end{itemize}
\end{enumerate}

\section{Example Ship Handout} \label{example_ship}

\begin{itemize}[noitemsep]
    \item Auxiliary cargo or support vessel (e.g., OSV)
    \item Approximately 90\,m LOA, 16\,m beam, and 5{,}000 gross tons
    \item Twin-screw diesel propulsion with controllable pitch propellers (CPP), independently controlled
    \item Propulsion control available remotely (bridge), from the engine control room, or locally
    \item Electrically driven bow thruster
    \item Twin hydraulic rudders
    \item Dual redundant steering pumps with manual operation capability
    \item Steering control available remotely (bridge), from the steering control room, or locally
    \item Integrated Bridge System (IBS) with Machinery Control and Monitoring System (MCMS)
    \item Dual redundant Ethernet network
    \item Two independent, integrated ECDIS units, configured and maintained in accordance with IMO standards
    \item No paper charts
    \item Two independent GNSS receivers (multiple constellations with DGPS)
    \item Dual radar systems
    \item Gyrocompass repeater with two alidades
    \item Secondary transmitting magnetic compass
    \item AIS Class~A, integrated with both multifunction displays (MFDs)
    \item Autopilot capable of heading hold and track control
    \item Speed log
    \item VHF and HF radio communications
\end{itemize}

\section{Supplementary Figures}
\label{app:figures}

\begin{figure}[ht]
    \centering
    \includegraphics[width=\columnwidth]{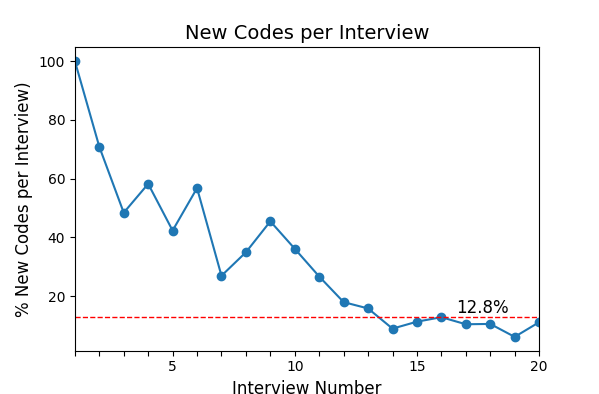}
    \caption{Proportion of new codes per interview.}
    \label{fig:sat_figure}
    \Description[Saturation Point Figure]{A line plot showing the percent of new codes per interview (y-axis) by interview number (x-axis). The plot trends downward, starting at 100 percent at interview 1 and falling to 12.8\% by interview 14. The percent of new codes does not rise above 12.8 percent after interview 14.}
\end{figure}

\end{document}